\documentclass[a4paper,12pt]{article}
\pdfoutput=1
\usepackage{jheppub}
\usepackage{slashed}
\usepackage[dvipsnames,table]{xcolor}
\usepackage{hyperref} 
\usepackage{xspace}
\usepackage[tight]{subfigure}
\usepackage{amsmath}
\usepackage{rotating}
\usepackage{amssymb}
\usepackage{amsfonts}
\usepackage{mathrsfs}
\usepackage{comment}
\usepackage{afterpage}
\usepackage{verbatim}
\usepackage{tabularray}
\UseTblrLibrary{booktabs}
\UseTblrLibrary{siunitx}
\UseTblrLibrary{functional}
\newcommand{\bftab}{\fontseries{b}\selectfont} 
\usepackage{array}
\usepackage[title,titletoc]{appendix}
\usepackage{tabularx}
\usepackage{mathtools}
\usepackage{bbm}
\usepackage{todonotes}
\usepackage{csquotes}
\usepackage{siunitx}
\usepackage[shortcuts]{extdash}
\newcommand{\GeV}{\ensuremath{\,\text{GeV}}\xspace}
\newcommand{\MeV}{\ensuremath{\,\text{MeV}}\xspace}
\newcommand{\Sherpa}{{\sc{Sherpa}}\xspace}
\newcommand{\Pepper}{{\sc{Pepper}}\xspace}
\newcommand{\Recola}{{\sc{Recola}}\xspace}
\newcommand{\OpenLoops}{{\sc{OpenLoops}}\xspace}
\newcommand{\MadNIS}{{\sc{MadNIS}}\xspace}
\newcommand{\MadLoop}{{\sc{MadLoop}}\xspace}
\newcommand{\MGaMCNLO}{{\sc{MadGraph5\_aMC@NLO}}\xspace}
\newcommand{\Stripper}{{\sc{Stripper}}\xspace}
\newcommand{\Vegas}{{\sc{Vegas}}\xspace}

\DeclarePairedDelimiterX{\infdivx}[2]{(}{)}{%
  #1\;\delimsize\|\;#2%
}

\renewcommand{\vec}[1]{\mathbf{#1}}

\DeclareMathOperator{\divergence}{div}
\newcommand{\order}[1]{\ensuremath{\mathop{}\mathopen{}\mathcal{O}\mathopen{}\left(#1\right)}}
\DeclarePairedDelimiter\abs{\lvert}{\rvert}

\newcommand{\dif}{\mathop{}\!\mathrm{d}}

\makeatletter
\newcommand{\spx}[1]{%
  \if\relax\detokenize{#1}\relax
    \expandafter\@gobble
  \else
    \expandafter\@firstofone
  \fi
  {^{#1}}%
}
\makeatother

\newcommand{\od}[3][]{\frac{\dif\spx{#1}#2}{\dif#3\spx{#1}}}

\let\dd\dif

\title{Sampling NNLO QCD phase space with normalizing flows}
\author[a,b]{Timo Janßen,}
\author[c]{Rene Poncelet,}
\author[b]{Steffen Schumann}
\affiliation[a]{Campus-Institut Data Science, Georg-August-Universit\"at G\"ottingen, Goldschmidtstra\ss{}e 1, 37077 G\"ottingen,  Germany}
\affiliation[b]{Institut für Theoretische Physik, Georg-August-Universit\"at G\"ottingen, Friedrich-Hund-Platz 1, 37077 G\"ottingen, Germany}
\affiliation[c]{Institute of Nuclear Physics, ul. Radzikowskiego 152, 31--342 Krakow, Poland}
\emailAdd{timo.janssen@theorie.physik.uni-goettingen.de}
\emailAdd{rene.poncelet@ifj.edu.pl}
\emailAdd{steffen.schumann@phys.uni-goettingen.de}

\abstract{
We showcase the application of neural importance sampling for the evaluation of NNLO QCD 
scattering cross sections. We consider Normalizing Flows in the form of discrete Coupling Layers and time continuous flows for the integration of the various cross-section contributions when using the sector-improved residue subtraction scheme. We thereby consider the stratification of the integrands into their positive and negative contributions, and separately optimize the phase-space sampler. We exemplify the novel methods for the case of gluonic top-quark pair production at the LHC at NNLO QCD accuracy. We find significant gains with respect to the current default methods used in STRIPPER in terms of reduced cross-section variances and increased unweighting efficiencies. In turn, the computational costs for evaluations of the integrand needed to achieve a certain statistical uncertainty for the cross section can be reduced by a factor 8.}

\preprint{IFJPAN-IV-2025-11, MCNET-25-11, COMETA-2025-22}

\begin{document}
\strut\hfill
\maketitle

\section{Introduction}\label{sec:intro}

Fixed-order perturbative calculations provide unique means to address scattering processes at high energy scales as probed in particle colliders. In particular when considering hadronic initial states, higher-order QCD predictions are vital for the success of experiments aiming to scrutinize our understanding of fundamental-particle interactions. Such fixed-order calculations require the evaluation of squared transition matrix elements in dependence on the initial- and final-state particle momenta. Besides the construction of compact expressions for the matrix elements, the evaluation of physical cross sections requires the efficient generation of phase-space configurations for the external states. The computation of tree-level cross sections has been fully automated in modern event-generator programs even for high-multiplicity processes~\cite{Kanaki:2000ey,Maltoni:2002qb,Krauss:2001iv,Kilian:2007gr,Gleisberg:2008fv}. While such leading-order (LO) squared matrix elements are strictly positive, this is no longer guaranteed at higher orders, e.g.\ due to interference contributions of Born and loop diagrams. Next-to-leading order (NLO) calculations can also be considered fully automated for Standard Model processes, thanks to dedicated loop-amplitude generators such as \OpenLoops~\cite{Cascioli:2011va,Buccioni:2019sur}, \Recola~\cite{Actis:2016mpe,Denner:2017wsf} or \MadLoop~\cite{Hirschi:2011pa}. Another vital ingredient for numerical evaluations of NLO cross sections are infrared subtraction methods such as Frixione--Kunszt--Signer (FKS)~\cite{Frixione:1995ms} and Catani--Seymour (CS) subtraction~\cite{Catani:1996vz,Catani:2002hc}. These accomplish the separate cancellation of soft- and/or collinear singularities in the real-emission and virtual corrections, thereby rendering them finite, opening the possibility to integrate both parts numerically. Efficient phase-space sampling for NLO calculations demands suitable treatment of real-emission configurations, probing both, hard and well separated emissions, as well as rather soft and collinear kinematics.   

Moving to next-to-next-to-leading order (NNLO) introduces further challenges. On the one hand, the computation of the required two-loop matrix elements is more complicated since no numerical method has been developed yet, necessitating (semi\mbox{-})analytic derivations on a case-by-case basis. For example, state-of-the-art computations derive amplitudes with two-to-three kinematics~\cite{Chawdhry:2019bji, Kallweit:2020gcp, Chawdhry:2021hkp, Czakon:2021mjy, Chen:2022ktf, Hartanto:2022qhh, Badger:2023mgf, Buccioni:2025bkl}. The second bottleneck arises from the organization of the cancellation of the infrared divergences in terms of subtraction or slicing methods. At present, several approaches are being explored. Each presents unique advantages and challenges and differs in generality and efficiency. Slicing methods like $q_T$ \cite{Catani:2007vq, Bonciani:2015sha} and N-jettiness \cite{Gaunt:2015pea, Boughezal:2015dva, Boughezal:2016zws, Moult:2016fqy, Moult:2017jsg, Ebert:2018lzn, Boughezal:2018mvf, Boughezal:2019ggi} slicing are easy to implement but typically rather inefficient. The slicing parameter also might restrict the type of processes the scheme can be applied for, like $q_T$\mbox{-}slicing, which works only for colour-singlet or massive final-state particles but not for jet production. N\mbox{-}jettiness slicing gets increasingly challenging if more than three coloured particles are involved due to the complex colour structures. NNLO subtraction methods follow the paradigm of the successful FKS and CS NLO schemes and implement local subtraction terms. Roughly speaking, the various approaches differ in three fundamental properties: identification of the singular integrable limits and construction of suitable subtraction terms, phase-space mappings, and the treatment of integrated subtraction terms (numerical or analytical). One can broadly separate FKS\mbox{-}like (the sector-improved residue \cite{Czakon:2010td, Czakon:2011ve, Czakon:2014oma, Czakon:2019tmo}, nested soft-collinear \cite{Caola:2017dug, Caola:2019nzf, Caola:2019pfz, Bargiela:2023npj, Devoto:2023rpv}, local analytic \cite{Magnea:2018hab, Bertolotti:2022aih, Magnea:2020trj} subtraction schemes), CS\mbox{-}like (antenna \cite{Gehrmann-DeRidder:2005btv, Gehrmann-DeRidder:2005svg, Gehrmann-DeRidder:2005alt, Daleo:2006xa, Daleo:2009yj, Boughezal:2010mc, Gehrmann:2011wi, Gehrmann-DeRidder:2012too, Currie:2013vh, Bernreuther:2011jt, Bernreuther:2013uma, Abelof:2011jv, Abelof:2012he, Abelof:2014fza, Currie:2018oxh, Gehrmann:2023dxm, Braun-White:2024ojt, Bonino:2024adk, Fox:2024bfp}, CoLoRfulNNLO \cite{Somogyi:2005xz, Somogyi:2006da, Somogyi:2006db, Aglietti:2008fe, DelDuca:2013kw}) and projection-to-Born schemes \cite{Cacciari:2015jma, Currie:2018fgr}.

For the feasibility of NNLO QCD calculations, the numerical efficiency of the employed subtraction scheme is a critical aspect. The typical computational costs are $\order{10^3}$ CPUh for simple colourless final states. If the final state is more complicated, particularly if it involves massless coloured particles, the resource demand increases. For example, an NNLO QCD computation for the hadronic production of vector-boson plus jet final states is already of $\order{10^4}$ CPUh, a di-jet computation typically required $\order{10^5}$ and three-jet calculations $\order{10^6}$ CPUh. These estimates apply for typical LHC fiducial phase spaces and single-differential observables. If higher-dimensional distributions or particular corners of phase space are considered, the costs might be even higher. The details depend on the scheme used in the computation, but usually, the main bottleneck turns out to be the phase-space integration of the double real-emission contribution and the numerical cancellation between various (integrated) subtraction terms and between different contributions, which require a very small residual statistical uncertainty on the individual components.

Accordingly, it is desirable to increase the efficiency of such numerical computations. Two avenues can be explored to achieve this: 1) Modifying the subtraction scheme in order to minimize the cancellations between contributions or 2) improving the numerical-integration techniques. In this work, we explore the second approach motivated by the fact that many NNLO QCD calculational frameworks use relatively simple techniques for numerical integration. To better adapt to the given integrand, it is common to use techniques like integration-variable re-mappings (importance sampling) and multi-channel phase-space parameterizations. However, the standard method to perform phase-space integration is to fall back to rather simple adaptive Monte Carlo techniques such as \textsc{Vegas}~\cite{Lepage:1977sw}.

The rapid advances of modern machine-learning techniques have sparked renewed interest and research activities in phase-space integration and sampling methods for high-energy physics event generators~\cite{Butter:2022rso}. In particular, this includes the development of Neural Importance Sampling methods~\cite{Bothmann:2020ywa,Gao:2020zvv} that employ trainable Normalizing-Flow transformations~\cite{tabak:2010,Tabak:2013cnz,pmlr-v37-rezende15,papamakarios:2021} to optimize the density of phase-space integration variables. Such methods have been developed in the context of the \Sherpa event generator~\cite{Sherpa:2024mfk,Sherpa:2019gpd}, the \MadNIS approach~\cite{Heimel:2022wyj,Heimel:2023ngj,Heimel:2024wph} for \MGaMCNLO~\cite{Alwall:2011uj,Alwall:2014hca}, see also Refs.~\cite{Gao:2020vdv,Verheyen:2022tov,Deutschmann:2024lml,Kofler:2024efb,ElBaz:2025qjp}. In essence, these methods offer potential to reduce the variance of cross-section integral estimates by reducing the spread of event weights. This will typically also yield improved unweighting efficiencies for the generation of unit-weight events as they are customarily used in experimental applications of LO and NLO accurate theory predictions. Recently, there have been first attempts to produce (partially) unweighted events also for NNLO QCD predictions~\cite{Czakon:2023hls}. For novel approaches to accelerate event unweighting, i.e.\ rejection sampling, through the use of neural-network matrix-element surrogates, see Refs.~\cite{Danziger:2021eeg,Janssen:2023ahv}.

In this publication, we pioneer the application of Neural Importance Sampling for phase-space integration in a complete NNLO QCD computation\footnote{Applications of Normalizing Flows for solving Feynman-parameter integrals, as they appear in higher-order perturbative calculations, have been presented in~\cite{Winterhalder:2021ngy,Jinno:2022sbr}.}. As an initial example, we consider the fully differential hadronic production of top-quark pairs in the \Stripper framework~\cite{Czakon:2015owf}, based on sector improved residue subtraction. To this end, we train and apply dedicated Normalizing-Flow samplers for the various components of the NNLO calculation, including the Born, single and double real-emission corrections, as well as one- and two-loop contributions. We here consider splitting the target distribution in positive and negative contributions and separately optimize samplers for the two pieces, resulting in a significant cross-section variance reduction and accordingly the computing resources needed to arrive at an accuracy target. For the Normalizing-Flow transformations we consider Coupling-Layer architectures~\cite{Dinh:2014} as well as Continuous Normalizing Flows~\cite{Chen:2018}. For the latter, we build on the experience and implementation from Ref.~\cite{janssen2025}, where orders of magnitude improvements in unweighting efficiency were achieved for the LO case using Flow Matching~\cite{lipman2023flow,tong2024improving} in conjunction with the portable event generator \Pepper~\cite{Bothmann:2023gew}. The corresponding results get compared to what is obtained with variable re-mappings based on \Vegas, what closely resembles the default approach used in \Stripper~\cite{vanHameren:2007pt}.

The article is structured as follows: In Sec.~\ref{sec:pssampling} we introduce the basic concepts of phase-space integration and sampling, including the generation of unweighted samples. We furthermore introduce the \Stripper approach to perform NNLO QCD calculations, thereby focusing on the structure of the contributing phase-space integrals. In Sec.~\ref{sec:flows} we briefly introduce Coupling-Layer and Continuous Normalizing Flows, describe our training approach, and how discrete integration variables, e.g.\ corresponding to helicity states of external particles, can be incorporated in conditional flows. Ultimately, in Sec.~\ref{sec:NNLO_results} we present the setups of our numerical calculations and the results obtained for gluonic top-quark pair production at NNLO QCD. Our conclusions and a brief outlook are compiled in Sec.~\ref{sec:conclusions}. 
 
\section{Phase-space sampling for scattering processes}
\label{sec:pssampling}

In this section we briefly recapitulate basic elements of Monte Carlo integration and importance sampling specifically. In particular we consider integrands of non-definite sign and study the impact of their stratification into positive and negative contributions. Further, in Sec.~\ref{subsec:integrands} we introduce the \Stripper approach to NNLO QCD calculations and thereby put the focus on the appearing phase-space integrals and their parameterization. 

\subsection{Phase-space integration basics}\label{sec:prelim1}

We consider a generic integration problem over some $d$-dimensional hypercube $\Omega$ of unit volume with a piece-wise Riemann integrable integrand $f:\Omega\to \mathbbm{R},\vec{x}\mapsto f(\vec{x})$:
\begin{equation}
    I = \int_{\vec{x} \in \Omega} \dd \vec{x} f(\vec{x})\,.
    \label{eq:integral}
\end{equation}
The Monte Carlo estimate for the integral and the associated one-sigma statistical error for $N$ independent uniformly distributed points $\vec{x}_i \in \Omega$ read
\begin{equation}\label{eq:integral_estimate}
    \hat{I} = \frac{1}{N} \sum_{i=1}^N f(\vec{x}_i) \;, \quad \delta \hat{I} = \sqrt{\frac{1}{N-1} \left(\frac{1}{N} \sum_{i=1}^N f^2(\vec{x}_i) - \hat{I}^2\right)}\;.
\end{equation}
To reduce the variance of the integral estimate, the density of sampling points can be adjusted to better resemble the target distribution $f(\vec{x})$, known as \emph{importance sampling}~\cite{kroese2013handbook}. To this end, a bijective mapping $\vec{H}: \Omega \to \Omega, \vec{x} \mapsto \vec{H}(\vec{x})$ can be considered, resulting in 
\begin{equation}\label{eq:ISdef}
    I = \int_{\vec{H}(\vec{x}) \in \Omega} \dd \vec{H} \,\frac{f(\vec{x})}{h(\vec{x})}\,,
\end{equation}
with the density $h:\Omega\to \mathbbm{R},\vec{x}\mapsto h(\vec{x})$ given by the Jacobian determinant
\begin{equation}
    h(\vec{x}) = \left|\det\left(\frac{\partial \vec{H}(\vec{x})}{\partial \vec{x}}\right)\right|\,. 
\end{equation}
The density function $h(\vec{x})$ might thereby depend on a set of parameters that can be adjusted in an initial optimization phase to optimally map the target distribution $f(\vec{x})$. However, in order to be able to efficiently generate phase-space points distributed according to $h(\vec{x})$, typically the \emph{inverse transform method}~\cite{Devroye:1986:NUR,kroese2013handbook} is used. This, however, requires an easily integrable form for $h(\vec{x})$, with an invertible cumulative distribution, what might limit the expressivity and thus the performance of the algorithm.

Should the integrand not be strictly positive, during this initial phase typically its modulus is considered for optimizing the sampling distribution $h(\vec{x})$. However, in the actual integration phase the sign of the target is kept and propagated accordingly. In this situation there exists a lower limit for the variance that can be derived from the case that $|f(x)|/h(x) = w = \text{const}$. In this extreme case, the mean is given by
\begin{equation}\label{eq:alpha_introduced}
    \hat{I} = w \frac{N_+ - N_-}{N} \equiv w (2\alpha-1)\,,
\end{equation}
where $\alpha = N_+/N$, with $N_{\pm}$ denoting the number of points ending up in the positive or negative region of phase space, respectively. For the variance we find
\begin{equation}
    \operatorname{Var}(\hat{I}) = w^2 - w^2(2\alpha-1)^2 = w^2(4\alpha(1-\alpha))\,.
\end{equation}
The relative uncertainty on the integral is then given by
\begin{equation}\label{eq:minerr_toy}
    \frac{\delta\hat{I}}{\hat{I}} = \frac{1}{\sqrt{N-1}} \frac{\sqrt{\alpha(1-\alpha)}}{\alpha-\frac{1}{2}}\,.
\end{equation}
For $\alpha<1$ this yields a non-vanishing lower bound for the relative error, independent of how well $h(x)$ approximates $f(x)$. The limit on the error $\delta\hat{I}$ we denote by $\delta_{\rm opt}$.

\subsubsection*{Stratification for signed integrands}
A possible alternative treatment, that we will explore in this paper, is to actually split the integrand into its positive and negative contributions, i.e.\ 
\begin{equation}
f(\vec{x})=f_+(\vec{x})+f_-(\vec{x})\,,\quad \text{with}\quad f_\pm(\vec{x})=\Theta(\pm f(\vec{x})) f(\vec{x})\,, 
\end{equation}
and to separately adapt suitable densities $h_\pm(\vec{x})$, with
$h_\pm:\Omega \to \mathbbm{R}, \vec{x}\mapsto h_\pm(\vec{x})$. We use \(\Theta\) to denote the Heaviside step function. 
Accordingly, the desired integral is given by
\begin{equation}\label{eq:fsplit}
    I = \int_{\vec{H}_+(\vec{x})\in\Omega} \dd \vec{H}_+ \,\frac{f_+(\vec{x})}{h_+(\vec{x})}+ \int_{\vec{H}_-(\vec{x})\in\Omega} \dd \vec{H}_- \,\frac{f_-(\vec{x})}{h_-(\vec{x})}\,.
\end{equation}
This can be regarded as an instance of \emph{stratified sampling}~\cite{kroese2013handbook} with two strata. In this case, the Monte Carlo estimate of the integral becomes
\begin{equation}
    \hat{I}_{\text{strat}} = \hat{I}_+ + \hat{I}_- = \frac{1}{N_+} \sum_{i=1}^{N_+} \frac{f_+(\vec{x}_i)}{h_+(\vec{x}_i)} + \frac{1}{N_-} \sum_{i=1}^{N_-} \frac{f_-(\vec{x}_i)}{h_-(\vec{x}_i)} \,.
\end{equation}
In contrast to above, there is no lower limit on the variance. If \(h_\pm\) approximate their respective targets \(f_\pm\) perfectly, the integral can be determined with vanishing error, i.e.\ $\delta_{\text{opt}}=0$. In the (realistic) case of finite variance, we need to consider the sum of the two variances weighted by their respective relative fractions, resulting in
\begin{align}
    \delta \hat{I}_{\text{strat}}             &= \sqrt{\frac{1}{N-1}\Biggl[\frac{N}{N_+} \operatorname{Var}\bigl(\hat{I}_+\bigr) + \frac{N}{N_-} \operatorname{Var}\bigl(\hat{I}_-\bigr)\Biggr]} \quad \text{with} \\
    \operatorname{Var}\bigl(\hat{I}_\pm\bigr) &= \frac{1}{N_\pm} \sum_{i=1}^{N_\pm} \left(\frac{f_\pm(\vec{x}_i)}{h_\pm(\vec{x}_i)}\right)^2 - \hat{I}_\pm^2 \,,
\end{align}
where we used \(N = N_+ + N_-\). This error can be smaller than the one using a single \(h(\vec{x})\) optimized on \(\abs{f(\vec{x})}\), provided that the individual variances \(\operatorname{Var}\bigl(\hat{I}_\pm\bigr)\) are small enough. This requires not only that the densities \(h_\pm(\vec{x})\) adapt well to the functional shape of \(f_\pm(\vec{x})\) but also that they efficiently sample within the volumes \(\Omega_\pm \subset \Omega\) where the target \(f_\pm(\vec{x})\) is non-zero. In terms of the latter property, Normalizing Flows should have an advantage over \Vegas, as they are not limited to a factorized approach.

\subsubsection*{Unweighted event generation}
In order to generate unit-weight events from the target function, a \emph{rejection-sampling algorithm}~\cite{vonNeumann1951} is employed. An event with weight $w_i=f(\vec{x}_i)$, i.e.\ $w_i=f(\vec{x}_i)/h(\vec{x}_i)$ when sampling with a non-uniform density, gets accepted with probability 
\begin{equation}
p_i=\frac{w_i}{w_\text{max}}\,,
\end{equation}
with $w_\text{max}$ a pre-determined maximal weight. Accordingly, the efficiency of the unweighting procedure is determined by
\begin{equation}\label{eq:unweighting_efficiency}
    \epsilon_\text{uw} \equiv \frac{\langle w\rangle}{w_\text{max}}\,,
\end{equation}
with $\langle w \rangle$ the average event weight in a given sample. The inverse $1/\epsilon_\text{uw}$ thereby determines the average number of target-function evaluations needed to generate a single unit-weight event. Note that for not strictly positive target functions the unit weights carry a sign. For discussions on the sources of negative event weights in high-energy event simulations at NLO QCD accuracy, their impact on the statistical properties of event samples, and ideas how to mitigate their contribution, see for example Refs.~\cite{Frederix:2020trv,Andersen:2020sjs,Danziger:2021xvr,Frederix:2023hom}.

To soften the dependence of the unweighting efficiency on rare outlier events featuring very large weights, the naively obtained maximum can systematically be reduced, e.g.\ by removing the largest weights up to a certain percentile of the total integral~\cite{Gao:2020zvv,Danziger:2021eeg}. Correspondingly, events whose weight $w_i$ exceeds the reduced maximum will pass the unweighting step and be assigned an overweight given by 
\begin{equation}\label{eq:overweight}
w^\text{overweight}_i=\text{sgn}(w_i)\left|\frac{w_i}{w^\text{reduced}_\text{max}}\right|\,.
\end{equation}

\subsection{Integrands in the sector-improved residue subtraction scheme}\label{subsec:integrands}
In this section we briefly review the anatomy of NNLO QCD calculations based on the sector-improved residue subtraction scheme as implemented in the \Stripper framework. We thereby focus on the structure of the appearing phase-space integrals to compute (differential) cross sections. Scattering cross sections are obtained upon integration of transition matrix elements over the momenta of external particles. The crucial point is that eventually, employing a suitable parameterization, the appearing integrals can be related to integrals over a unit hypercube in the sense of Eq.~\eqref{eq:integral}.

The construction starts from the expression of the (differential) hadronic cross section of producing a final state $X$ in collinear factorization, given by 
\begin{equation}
  \sigma(h(P_1) h(P_2) \to X) = \sum_{ab} \iint_0^1 \dd x_1 \dd x_2 \phi_a(x_1)\phi_b(x_2) \hat{\sigma}_{ab}(x_1P_1,x_2P_2)\,,
  \label{eq:tot_xsec}
\end{equation}
where we have suppressed the dependence on the factorization and renormalization scales. Here the sum of $a$ ($b$) runs over all massless parton flavours, i.e.\ gluons $g$ and quarks $q\in\{u,d,c,s,b\}$, as well as their anti-particles $\bar{q}$. The functions $\phi_a$ denote the corresponding renormalized Parton-Distribution-Functions (PDFs). $\hat{\sigma}_{ab}$ represents the partonic cross section, whose expansion in the strong coupling constant $\alpha_s$ up to NNLO can be written as
\begin{equation}
  \hat{\sigma}_{ab} = \hat{\sigma}^{(0)}_{ab} + \hat{\sigma}^{(1)}_{ab} + \hat{\sigma}^{(2)}_{ab} + \order{\alpha_s^{n_{(0)}}\alpha_s^{3}}\,.
  \label{eq:partonic_xsec_exp}
\end{equation}
Powers of the strong coupling (including the corresponding tree-level coupling order $n_{(0)}$) are thereby absorbed in the expansion coefficients. The leading-order or Born-level contribution reads
\begin{equation}
    \hat{\sigma}^{(0)}_{ab} = \hat{\sigma}^{\text{B}}_{ab}\,.
\end{equation}
At NLO QCD there appear virtual ($\hat{\sigma}_{ab}^{\text{V}}$), real emission ($\hat{\sigma}_{ab}^{\text{R}}$), and, for hadronic initial states, collinear PDF counterterm ($\hat{\sigma}_{ab}^{\text{C}}$) contributions. Thus, the complete NLO correction reads
\begin{equation}\label{eq:NLOcontribs}
    \hat{\sigma}^{(1)}_{ab} = \hat{\sigma}^{\text{R}}_{ab} + \hat{\sigma}^{\text{V}}_{ab}  + \hat{\sigma}^{\text{C}}_{ab}\,.
\end{equation}
At NNLO QCD double-virtual ($\hat{\sigma}_{ab}^{\text{VV}}$), double-real ($\hat{\sigma}_{ab}^{\text{RR}}$), and real-virtual ($\hat{\sigma}_{ab}^{\text{RV}}$) contributions have to be taken into account, potentially together with collinear PDF counterterms ($\hat{\sigma}_{ab}^{\text{C1}}$, $\hat{\sigma}_{ab}^{\text{C2}}$). Accordingly, the NNLO QCD correction can be written as:
\begin{equation}\label{eq:NNLOcontribs}
    \hat{\sigma}^{(2)}_{ab} = \hat{\sigma}^{\text{RR}}_{ab} + \hat{\sigma}^{\text{RV}}_{ab}  + \hat{\sigma}^{\text{VV}}_{ab} + \hat{\sigma}^{\text{C1}}_{ab} + \hat{\sigma}^{\text{C2}}_{ab} \;.
\end{equation}
For each contribution $i$ we denote $\sigma^{i}$, i.e.\ with the PDF convolution in Eq.~\eqref{eq:tot_xsec} performed, as the hadronic contribution to the cross section.

Besides the various contributions from real and virtual corrections, the partonic cross section is furthermore composed of different partonic final states that need to be summed over. We absorb all these contributions in a generic index $j \in \mathcal{C}^{i}_{ab}(ab \to X)$, where $\mathcal{C}^{i}_{ab}(ab \to X)$ is the set of indices required for a given contribution $i$ and Born process $ab \to X$, accordingly
\begin{equation}
    \hat{\sigma}^{i}_{ab} = \sum_{j \in \mathcal{C}^{i}_{ab}(ab \to X)} \hat{\sigma}^{i}_{ab,j}\,.
\end{equation}
The individual contributions can be cast in the form
\begin{equation}
  \hat{\sigma}^{i}_{ab,j} = N_j \int\dd \Phi_{j} \mathcal{F}_{ab,j}(\Phi_j)\,,
  \label{eq:generic_con}
\end{equation}
with $\Phi_j$ the Lorentz-invariant phase space\footnote{This phase-space measure might include the integration over convolution variables in case of collinear factorization counterterms.}, and $\mathcal{F}_{ab,j}$ a function of the kinematics, representing transition matrix elements and subtraction terms, as well as phase-space cuts.

Consider, as an example, the LO case: the function $\mathcal{F}_{ab,j}$ is simply a matrix element times a measurement function ($F$), used to define infrared-safe observables and the fiducial phase space:
\begin{equation}
\mathcal{F}_{ab,j} = |\mathcal{M}_{ab,j} (\Phi_{j})|^2 F(\Phi_{j})\,.
\end{equation}
At higher orders, real-radiation contributions require the subtraction of infrared limits and their combinations with virtual infrared divergences. In the \Stripper scheme, we introduce for this purpose selector functions, see \cite{Czakon:2014oma}. In the case of a single-real emission, where $j$ represents an $n+1$ partonic process, we have:
\begin{equation}
\hat{\sigma}_{ab}^{(1)} \ni \int \dd \Phi_j \sum_{kl} \mathcal{S}_{kl} |\mathcal{M}_{j} (\Phi_{j})|^2F(\Phi_{j})\,.
\end{equation}
Selector functions have the property that they sum up to unity:
\begin{equation}
\sum_{kl} \mathcal{S}_{kl} = 1\,,
\end{equation}
and that they vanish sufficiently fast in infrared limits not given by the soft limit of parton $k$ and the collinear limit $k \to l$. This property allows one to explicitly parameterize these limits in terms of variables $\xi$ (energy of $k$) and $\eta$ (angle between $k$ and $l$). Details on the phase-space parameterization can be found in Ref.~\cite{Czakon:2019tmo}. After the transformation and parameterization of the rest of the phase space, we find integrals of the form (now for each sector individually):
\begin{equation}
\hat{\sigma}_{ab}^{(1)} \ni \iint_0^1 \frac{\dd \xi}{\xi^{1-2\varepsilon}}
                                \frac{\dd \eta}{\eta^{1-\varepsilon}}
           \underbrace{\int \dd \tilde{\Phi}_j \mathcal{S}_{kl}
                       |\mathcal{M}_j (\Phi_{j})|^2F(\Phi_{j})}_{
             \text{finite} \;\equiv\; f^j_{kl}(\xi,\eta)}\,.
\end{equation}
This form makes it explicit that the only remaining singularities are given by the poles in $\xi$ and $\eta$. The plus distribution can be employed to extract these poles:
\begin{equation}
\begin{split}
\hat{\sigma}_{ab}^{(1)} \ni & \iint_0^1 \dd \xi \dd \eta \frac{f^j_{kl}(\xi,\eta)-f^j_{kl}(0,\eta)-f^j_{kl}(\xi,0)+f^j_{kl}(0,0)}{\xi^{1-2\varepsilon}\eta^{1-\varepsilon}}\\
    & +\frac{1}{2\varepsilon} \int_0^1 \dd \eta \frac{f^j_{kl}(0,\eta) - f^j_{kl}(0,0)}{\eta^{1-\varepsilon}}\\
    &  +\frac{1}{\varepsilon}  \int_0^1 \dd \xi  \frac{f^j_{kl}(\xi,0) - f^j_{kl}(0,0)}{\xi^{1-2\varepsilon}}\\
    &  +\frac{1}{2\varepsilon^2} f^j_{kl}(0,0)\,,
\end{split} \label{eq:sigma1R}
\end{equation} 
what yields a set of finite integrals that can be evaluated by numerical integration after expansion in $\varepsilon$. Integrals with poles cancel against virtual corrections, and only their $\varepsilon^0$ coefficient contributes to the cross section. The construction of the (integrated) subtraction terms is expressed through splitting and soft-functions and lower-order matrix elements as detailed in~\cite{Czakon:2014oma, Czakon:2019tmo}. In the light of numerical integration, it is important to note that integrals with poles (i.e.\ integrated subtraction terms) are much easier to evaluate and integrate than the actual subtracted finite term (the first line in Eq.~\eqref{eq:sigma1R}), since they only consist of lower-multiplicity matrix elements and any remaining subtraction is in variables that do not change the kinematics if the parameterization is suitably chosen \cite{Czakon:2019tmo}. 

A similar construction can be performed at NNLO QCD for the double-real-emission contribution~\cite{Czakon:2014oma, Czakon:2019tmo}. Focusing on the more challenging finite part, taking the limit $\varepsilon \to 0$ (which can be done as all divergencies are properly subtracted) and including also the parameterization of the finite integrals in terms of a hypercube, the integration problem can be reformulated as
\begin{equation}\label{eq:subtracted_contribution}
\hat{\sigma}_{ab}^{(i)} \ni \int_{[0,1]^{m}}  \dd^m \vec{\chi} \int_{[0,1]^{n}} \dd^{n} \vec{x} \frac{f^j_{\{k\}}(\vec{\chi},\vec{x})-\sum f^j_{\{k\}}(\vec{\chi}|_{\to 0},\vec{x})}{\prod^m_i \chi_i}\equiv \int_{[0,1]^{n+m}} \dd \vec{y} g^j_{\{k\}}\,.
\end{equation}
Here, the selector indices are shown in a compressed notation $\{k\}$ (each $\{k\}$ corresponds to a set of indices, $i,j$ for single-collinear, $i,j,k$ for triple-collinear and $i,j,k,l$ for double-collinear sectors). The total number of singular phase-space integrals $m$ is $2$ for one emission and $4$ for two emissions, respectively. The collection of subtraction terms is denoted by $\sum f^j_{\{k\}}(\vec{\chi}|_{\to 0},\vec{x})$, where the sum runs over all possibilities to distribute zeros in the $\chi$ vector. Formulated this way and including the PDF convolutions as well as the summation over all sectors and partonic channels, it is natural to view the final integration problem in a form which resembles Eq.~\eqref{eq:integral}:
\begin{equation}\label{eq:PSgenericStripper}
  \sigma^{i} = \sum_{ab} \sum_{j \in \mathcal{C}^{i}_{ab}(ab \to X)} \sum_{\{k\}} \int_{[0,1]^{n+2}} \dd x_1 \dd x_2 \dd \vec{y} g^j_{\{k\}}(\vec{y})\,.
\end{equation}
This procedure, as discussed in more detail in Ref.~\cite{Czakon:2019tmo}, splits the contributions appearing in Eqs.~\eqref{eq:NLOcontribs} and \eqref{eq:NNLOcontribs} further into pieces that can individually be numerically integrated. They are sorted into the following groups:
\begin{eqnarray}
    \sigma^{\text{R}} &=& \sigma^{\text{RF}} + \sigma^{\text{RU}}\,,\label{eq:sigmaR}\\
    \sigma^{\text{V}} &=& \sigma^{\text{VF}} + \sigma^{\text{VU}}\,,\\
    \sigma^{\text{RR}} &=& \sigma^{\text{RRF}} + \sigma^{\text{RRSU}}+ \sigma^{\text{RRDU}}\,,\\
    \sigma^{\text{RV}} &=& \sigma^{\text{RVF}} + \sigma^{\text{RVDU}}+ \sigma^{\text{RVFR}}\,,\\
    \sigma^{\text{VV}} &=& \sigma^{\text{VVF}} + \sigma^{\text{VVDU}}+ \sigma^{\text{VVFR}}\label{eq:sigmaVV}\,.
\end{eqnarray}
To increase the computational efficiency, a couple of optimizations are performed:
\begin{itemize}
    \item \textbf{Sector sampling} is used to evaluate the sum over sectors, i.e.\ all choices for $\{k\}$. This is realized by randomly choosing the sector 
    for a Monte Carlo point (based on some fixed and optimized weights) or as an additional one-dimensional integration.
    \item \textbf{Helicity sampling} is employed for real-emission corrections to reduce the computational time spent on matrix-element evaluations. The sectors identify a set of partons (e.g.\ $i$ and $j$ in the case of single-collinear sectors) for which the sum over helicities is performed explicitly:
    \begin{equation}
        \int_{[0,1]^n} \dd \vec{y} g^j_{\{k\}}(\vec{y}) = \sum_{h \in \text{hel}_{\{k\}}}\int_{[0,1]^n} \dd \vec{y} g^j_{\{k\},h}(\vec{y})\,.
        \label{eq:helicity_sampling}
    \end{equation}
    This sum can be taken out and performed via Monte Carlo methods similar to the sector sampling.
    \item \textbf{Mapping of PDF channels} can be used to reduce the sum over the 121 parton-flavour combinations in Eq.~\eqref{eq:tot_xsec} to a smaller set of generic processes. When there are two (or more) pairs of initial-state partons $(a,b)$ and $(a',b')$, where
    \begin{equation}
        \hat{\sigma}_{ab} (x_1P_1,x_2P_2) = \hat{\sigma}_{a'b'}(x_1P_1,x_2P_2)\,,
    \end{equation}
    the expression can be rewritten as
    \begin{equation}
       \sigma(h(P_1)h(P_2)\to X) = \sum_{c\in \mathcal{C}}\iint_0^1 \dd x_1 \dd x_2 \left(\sum_{(ab)\in c} \phi_a(x_1)\phi_b(x_2)\right) \hat{\sigma}_{ab}\,,
    \end{equation}
    i.e.\ the sum over PDFs is performed for each phase-space point. 
\end{itemize}

\noindent
Ultimately, we sequence the discrete selection choices to a single discrete index $l$, which represents a specific combination of $a,b,j,\{k\},h$ in Eq.~\eqref{eq:PSgenericStripper} such that:
\begin{equation}\label{eq:PSI_general}
      \sum_{ab} \sum_{j \in \mathcal{C}^{i}_{ab}(ab \to X)} \sum_{\{k\},h}\int_{[0,1]^n} \dd \vec{y} g^j_{\{k\},h}(\vec{y}) = \sum_l  \int_{[0,1]^n} \dd \vec{y} g_l(\vec{y})\,.
\end{equation}
The sum over $l$ can again be performed in a Monte Carlo fashion.

In addition to the numerical integration of the (differential) cross section, the samples can be (partially) unweighted as discussed at the end of Sec.~\ref{sec:prelim1}. For example, such an unweighting of NNLO QCD partonic events has been performed in the context of the HighTEA project \cite{Czakon:2023hls}. A separate maximum weight $w_{\rm max}$ for the positive and negative regions of the phase space is used for the HighTEA procedure, and a reduction of the maximum weight according to Eq.~\eqref{eq:overweight} is tuned to achieve an unweighting efficiency of $\sim 0.1\%$ for all contributions. The negative and positive regions of phase space are unweighted with their own respective maximum weight. The admixture of positive and negative events and different contributions is chosen to optimize the numerical integration uncertainty for some benchmark differential distributions for a given process. 

\section{Normalizing Flow improved sampling for NNLO QCD}\label{sec:flows}

In this section we describe how the mapping \(\vec{H}\), introduced in Sec.~\ref{sec:prelim1} for transporting a simple base density to a density that is closer to the target, can be realized by Normalizing Flows. Through specific architecture choices, explained below, Normalizing Flows can be designed in a way so that their Jacobian determinant is tractable without having to invert the neural networks. Therefore, neural networks of arbitrary complexity, trained by stochastic gradient descent, can be used. In Sec.~\ref{sec:coupling_layer_flows}, we describe flows based on Coupling Layers, which implement the flow as a chain of discrete steps. As an alternative approach, Sec.~\ref{sec:ode_flows} deals with Continuous Normalizing Flows, where the flow is regarded as a transformation continuous in time. The training of both kinds of flows, specific to our application, is described in Sec.~\ref{subsec:flow_training}. Lastly, we consider flows conditioned on discrete inputs in Sec.~\ref{subsec:conditional_flows}.

\subsection{Coupling Layer flows}\label{sec:coupling_layer_flows}

In general, Normalizing Flows provide a bijective mapping between two probability densities. Typically, they map from a simple base density, e.g.\ a Gaussian or uniform one, to a complex density that approximates a target. The key idea of Normalising Flows can best be understood from the change of variables formula for probability densities:
\begin{equation}\label{eq:change_of_variables}
    \vec{y}_1 = \phi(\vec{y}_0) \quad \Rightarrow \quad \log q(\vec{y}_1) = \log q_0(\vec{y}_0) -\log \biggl| \det \frac{\partial \phi}{\partial \vec{y}_0} \biggr| \,.
\end{equation}
A simple base density \(q_0\) can be transformed into a complicated density \(q\) by constructing a suitable flow \(\phi: \Omega \to \Omega\). The main difficulty is the computation of the Jacobian determinant \(\det \frac{\partial \phi}{\partial \vec{y}_0}\). In general, this has a cubic cost in \(n\), the number of dimensions of \(\vec{y}\). However, this can be reduced by constraining the flow, for which different approaches are possible. Here we consider Coupling Layers~\cite{Dinh:2014}, which split the input into two partitions \(A\) and \(B\) and transform them according to
\begin{equation}\label{eq:coupling_layer}
  \begin{split}
    \vec{y}_1^A &= \vec{y}_0^A \,, \\
    \vec{y}_1^B &= C \bigl( \vec{y}_0^B; m_\theta(\vec{y}_0^A) \bigr) \,,
  \end{split}
\end{equation}
where \(m_\theta\) is a neural network with trainable parameters \(\theta\) and the coupling transform \(C\) is an invertible function that transforms \(\vec{y}_0^B\) element-wise.

For spline-based flows~\cite{Mueller:2018}, we restrict the input domain of \(C\) to \([0,1]^{|B|}\), with \(|B|\) the cardinality of partition \(B\), and divide each dimension into \(K\) bins. In the form of a spline, piece-wise functions defined in each bin are joined into a continuous curve. Here we consider Neural Spline Flows~\cite{Durkan:2019}, which use rational-quadratic splines, whose parameters (bin widths, bin heights, derivatives at internal knots) are determined by the neural-network output \(m_\theta(\vec{y}_0^A)\).

As can be seen from Eq.~\eqref{eq:coupling_layer}, a single Coupling Layer transforms only a part of the input. To obtain an expressive transformation, several layers can be composed in a chain \(\phi = \phi_k \circ \cdots \circ \phi_1\) with permutations of the variables in-between. To capture all possible correlations, one needs at least \(2 \lceil \log_2 n \rceil\) coupling layers if \(n>5\) and \(n\) layers if \(n \leq 5\)~\cite{Gao:2020vdv}.

Coupling Layers based on spline transformations naturally live on the unit hypercube \(\Omega\). This means we can simply use a uniform base distribution with constant density \(q_0\) on \(\Omega\) and use the Coupling Layer flow to map it to a non-uniform distribution on \(\Omega\). According to Eq.~\eqref{eq:change_of_variables}, the model density is then given by the Jacobian determinant of the flow.

\subsection{Continuous Normalizing Flows}\label{sec:ode_flows}

An alternative to flows based on a chain of Coupling Layers is given by Continuous Normalizing Flows, which provide a flow \(\phi_t: [0, 1] \times \mathbbm{R}^n \to \mathbbm{R}^n\) continuous in a \enquote{time} parameter \(t \in [0, 1]\). The change of variable now reads
\begin{equation}
    \log q_t(\vec{y}_t) = \log q_0(\vec{y}_0) -\log \biggl| \det \frac{\partial \phi_t}{\partial \vec{y}_0} \biggr| \,,
\end{equation}
where \(q_t\) is a time-dependent probability density function. The flow itself is constructed implicitly via the ordinary differential equation (ODE)
\begin{equation}\label{eq:ode}
    \od{}{t} \phi_t(\vec{y}) = v_t(\phi_t(\vec{y}))\,,\quad\text{with}\quad 
    \phi_0(\vec{y})= \vec{y} \,,
\end{equation}
where \(v_t : \mathbbm{R}^n \to \mathbbm{R}^n\) is a time-dependent vector field. A neural network \(v_{t;\theta}(\vec{y})\), with trainable parameters \(\theta\), can be used to model the vector field~\cite{Chen:2018}. As the probability is locally conserved, this vector field needs to obey a continuity equation. 

To generate a point \(\vec{y}_1\) with the flow \(\phi_t\), we need to sample an initial point \(\vec{y}_0 \sim q_0\) and integrate the ODE \eqref{eq:ode} over time:
\begin{equation}\label{eq:cnf_map}
    \vec{y}_1 = \phi_1(\vec{y}_0) = \int_0^1 v_t(\phi_t(\vec{y}_0)) \dif t \,.
\end{equation}
In practice, this can be done approximately with a numerical ODE solver. If, vice versa, we want to evaluate the model probability density at a given point \(\vec{y}_1\), we need to solve the reversed ODE
\begin{equation}\label{eq:reverse_ode}
    \od{}{s} \begin{bmatrix}
           \phi_{1-s}(\vec{y}) \\
           f(1-s)
         \end{bmatrix} = \begin{bmatrix}
                           -v_{1-s}(\phi_{1-s}(\vec{y})) \\
                           - \divergence\bigl(v_{1-s}(\phi_{1-s}(\vec{y}))\bigr)
                          \end{bmatrix}
\end{equation}
for \(s \in [0, 1]\), with the initial conditions given by
\begin{equation}
    \begin{bmatrix}
        \phi_1(\vec{y}) \\
        f(1)
    \end{bmatrix} = \begin{bmatrix}
                           \vec{y}_1 \\
                           0
                          \end{bmatrix} \,.
\end{equation}
The solution for the model log probability is then found to be
\begin{equation}\label{eq:cnf_log_prob}
    \log q_1(\vec{y}_1) = \log q_0(\vec{y}_0) - f(0) \,.
\end{equation}
Eqs.~\eqref{eq:reverse_ode}--\eqref{eq:cnf_log_prob} follow from the continuity equation~\cite{lipman2023flow}.

In contrast to spline-based \textsc{Cl} flows, \textsc{Ode} flows are in general defined on \(\mathbbm{R}^n\) instead of the unit hypercube \(\Omega\). To deal with this, we use the sigmoid function \(\sigma_{\text{sigmoid}}: \mathbbm{R} \to \Omega, x \mapsto \sigma_{\text{sigmoid}}(x)=1/(1+e^{-x})\) and its inverse, the logit function, to bijectively map between the two domains. For the \textsc{Ode} flow, the base density \(q_0\) is a standard normal. To sample a point in \(\Omega\), we draw a point \(\vec{y}_0 \in \mathbbm{R}^n\) from \(q_0\), use the flow to map it to a point \(\vec{y}_1 \in \mathbbm{R}^n\) according to Eq.~\eqref{eq:cnf_map} and then map it to \(\sigma_{\text{sigmoid}}(\vec{y}_1) \in \Omega\). The overall density on \(\Omega\) is given by the base density \(q_0(\vec{y}_0)\) multiplied by the Jacobian determinants of the two maps. For training with data living on \(\Omega\), they need to be mapped to \(\mathbbm{R}^n\) first with the logit transform. Note that in order to avoid numerical problems at the edges of phase space, we use a simple scale and shift transformation, \(x \mapsto x \cdot (1-\epsilon) + \epsilon/2\), where \(\epsilon\) is a small number, so that the regions where the logit function is steepest are avoided. Throughout this work, we use \(\epsilon = \num{1e-6}\).

\subsection{Training of Normalizing Flows}\label{subsec:flow_training}

Since we can evaluate the model probability of a Normalizing Flow -- both for the discrete, Coupling Layer based, and the continuous flow -- we can train it by minimizing the Kullback--Leibler (KL) divergence~\cite{Kullback:1951zyt} between the target density, here denoted $p(\vec{y})$, and the model density $q_\theta(\vec{y})$, which in the long term is equivalent to maximum-likelihood estimation\footnote{The KL-divergence requires normalized densities as input. The fact that the normalization of the cross section is a priori unknown does not pose a problem in practice, only the derivative of the KL-divergence with respect to the model parameters is needed for optimization. For gradient-descent optimizers like Adam that dynamically set their step sizes the unknown normalization factor cancels out. See, for example, Ref.~\cite{Mueller:2018}.} The KL divergence is asymmetric in its arguments and in practice we prefer the forward version,
\begin{equation}\label{eq:forward_kl}
    D_{\text{KL}}(p \parallel q_\theta) = \int \dd \vec{y} p(\vec{y}) \log\biggl(\frac{p(\vec{y})}{q_\theta(\vec{y})}\biggr) = \mathbbm{E}_{p(\vec{y})}\biggl[ \log\biggl(\frac{p(\vec{y})}{q_\theta(\vec{y})}\biggr) \biggr] \,,
\end{equation}
due to its mode covering behaviour. This avoids severe local underestimation of the probability density, which would lead to huge importance weights \(w\) and therefore high variance of the integral estimate \(\hat{I}\), cf.\ Eq.~\eqref{eq:integral_estimate}, and low unweighting efficiency \(\epsilon_{\text{uw}}\), cf.\ Eq.~\eqref{eq:unweighting_efficiency}. However, Eq.~\eqref{eq:forward_kl} assumes that we sample \(\vec{y} \sim p(\vec{y})\). Since our integration method is based on importance sampling, we do not actually produce data from the target density \(p\). Instead, we sample \(\vec{y} \sim q_\theta(\vec{y})\) with weights \(w(\vec{y}) = p(\vec{y}) / q_\theta(\vec{y})\), where the weights must necessarily be taken into account in order to reproduce the correct target distribution. Generating unweighted data through a rejection-sampling step would be too inefficient for the training. However, as shown in Ref.~\cite{Mueller:2018}, we can find a suitable objective by looking at the gradient of \(D_{\text{KL}}\) with respect to the model parameters:
\begin{equation}
  \begin{split}
    \nabla_\theta D_{\text{KL}}(p \parallel q_\theta) &= \nabla_\theta \int \dd \vec{y} p(\vec{y}) \log\biggl(\frac{p(\vec{y})}{q_\theta(\vec{y})}\biggr) \\
    &= - \int \dd \vec{y} p(\vec{y}) \nabla_\theta \log q_\theta(\vec{y}) \\
    &= - \int \dd \vec{y} q_\theta(\vec{y}) \frac{p(\vec{y})}{q_\theta(\vec{y})} \nabla_\theta\log q_\theta(\vec{y}) \\
    &= \mathbbm{E}_{q_\theta(\vec{y})}\biggl[ -\frac{p(\vec{y})}{q_\theta(\vec{y})} \nabla_\theta \log q_\theta(\vec{y}) \biggr] \,.
  \end{split}
\end{equation}
This means we can train on weighted data by simply weighting the negative log-likelihood with the importance weight \(w(\vec{y}) = p(\vec{y}) / q_\theta(\vec{y})\) while ignoring its gradient in the optimization. The objective requires a tractable target density \(p(\vec{y})\) but does not require its gradient, which means the flow can be adapted to targets that cannot easily be made differentiable, like those in large existing code bases. 

Due to the appearance of the importance weights in the objective, we can expect slow convergence as long as \(q\) is far away from \(p\). For this reason, we employ an iterative training scheme in Sec.~\ref{sec:flows_for_nnlo}, where a flow trained on a limited set of data is used to produce the next set of training data with better distribution. This is similar to the approach proposed in Ref.~\cite{Akhound:2024}. However, there the authors use a diffusion model and add new data to a buffer in each iteration, while we use flow matching and replace the data in each iteration. There are also similarities to the buffered training used by \MadNIS~\cite{Heimel:2022wyj}. The main difference to our approach is that they perform the training in a single run with a buffer of weighted events that is continuously replaced. In contrast to that, we use a fixed set of training data during each iteration, which gets completely replaced for the next iteration. The advantage is that we can separate the production of the training data from the model training. This allows us to use the hardware efficiently, since \Stripper\ is running on CPUs while we train and run our models on GPUs. In each iteration, data is pushed to the GPU memory only once, minimizing overhead for input/output operations. Furthermore, it allows us to design a minimalistic interface to \Stripper, as it only needs to read in random numbers and Jacobians and write out weights. Disadvantages are the disk space required to store the event samples and the more complicated workflow due to the iterations. However, batch processing on a compute cluster simplifies the latter considerably.

For Continuous Normalizing Flows, the use of a numerical ODE solver makes training and evaluation of the density slow in comparison to Coupling Layer flows. This motivated the development of simulation-free training methods~\cite{lipman2023flow,albergo2023building,liu2022}, which do not require an explicit simulation during training. The idea is to instead directly match the vector field \(v_{t;\theta}(\vec{y})\) to a target vector field \(u_t(\vec{y})\) which generates the target density \(p_1(\vec{y})\) through a target probability density path \(p_t(\vec{y})\). This is intractable in general since \(u_t(\vec{y})\) is unknown. However, we can construct conditional probability paths and vector fields which are conditioned on individual samples from the target distribution and the base distribution. This leads to the Conditional Flow Matching objective~\cite{tong2024improving}
\begin{equation}\label{eq:cfm_objective}
    \mathcal{L}_{\text{CFM}}(\theta) = \mathbbm{E}_{t, q(\vec{z}), p_t(\vec{y}|\vec{z})} \|v_{t;\theta}(\vec{y}) - u_t(\vec{y}|\vec{z})\|^2 \,,
\end{equation}
where \(\vec{z}\) is a latent conditioning variable. Here, we use the independent coupling, where \(q(\vec{z}) = q(\vec{y}_0)q(\vec{y}_1)\) with \(\vec{y}_0\) and \(\vec{y}_1\) being samples from the base and the target distribution, respectively. The conditionals are Gaussian flows between \(\vec{y}_0\) and \(\vec{y}_1\) with standard deviation \(\sigma\), given by
\begin{align}
    p_t(\vec{y}|\vec{z}) &= \mathcal{N}(\vec{y} \,|\, t\vec{y}_1 + (1-t) \vec{y}_0, \sigma^2)\,, \label{eq:CFM_conditional_density} \\
    u_t(\vec{y}|\vec{z}) &= \vec{y}_1 - \vec{y}_0 \,.
\end{align}
The standard deviation \(\sigma\) serves as a hyperparameter that introduces noise into the training. Remarkably, minimizing Eq.~\eqref{eq:cfm_objective} is equivalent to a regression of \(v_t(\vec{y}; \theta)\) to the (unknown) marginal target vector field \(u_t(\vec{y})\), since it has the same gradients~\cite{lipman2023flow,tong2024improving}.

Similar to above, we have to account for the fact that in our application, \(\vec{y}_1\) is sampled from the flow instead of the target. We can easily correct for this by introducing the importance weight as a multiplicative factor into Eq.~\eqref{eq:cfm_objective}. This is a variant of the Energy Conditional Flow Matching objective proposed in Ref.~\cite{tong2024improving}. We then use the same iterative training scheme as introduced above for the Coupling Layer flow.

\subsection{Conditional Flows}\label{subsec:conditional_flows}

As described in Sec.~\ref{subsec:integrands}, phase-space sampling in \Stripper\ involves sampling over a discrete index \(l\), representing a combination of sector and helicity. This can be implemented by a categorical distribution, for which the probabilities of the categories can be adapted to data based on their contribution to the total cross section of the integral with index \(l\). At the same time, the amplitudes to be evaluated are conditioned on \(l\). Traditionally, the integrator (and its \Vegas\ mapping) is blind to this and only adapts to the marginal amplitudes. While it would be possible to adapt individual \Vegas\ mappings for each index \(l\), a poor convergence would be expected at least for those indices where few events are generated due to their low probabilities.

In contrast to \Vegas, Normalizing Flows can themselves be conditioned on \(l\) and are able to learn the correlations between different indices. This is particularly interesting in view of the fact that the amplitude distributions for different indicies are often quite similar to each other. Sometimes, they only differ in the marginal distribution of one parameter. Therefore, learning the conditional amplitudes can be expected to be more efficient with respect to the amount of training data than learning each marginal amplitude individually.

Technically, the conditional Normalizing Flow provides a conditional model density \(q_{\theta}(\vec{y}|l)\). For the Coupling Layer Flow, this means that each coupling transform C in Eq.~\eqref{eq:coupling_layer} has to be conditioned on \(l\). This can be achieved by conditioning the neural networks, such that \(m_\theta=m_\theta(\vec{y}_0^A | l)\). In consequence, the conditioning variable \(l\) is independently fed into each coupling layer. For Continuous Normalizing Flows, only the vector field is conditioned, such that \(v_{t;\theta} = v_{t;\theta}(\vec{y}|l)\). 

Since the index \(l\) is categorical, it needs to be transformed to a continuous representation before being processed by a neural network. We use embedding layers for this purpose. There are different possibilities to feed the representation into the network. For simplicity, we concatenate the representation of \(l\) and the other input variables and jointly feed them into the input layer.

\section{Application to top-quark pair production at NNLO QCD}\label{sec:NNLO_results}

To apply, validate and benchmark our novel integration methods for NNLO QCD calculations in the 
\Stripper approach, we consider hadronic top-quark pair production. To ease the 
calculations, we here focus on the purely gluonic channel, i.e.\ no contributions from light-quark 
flavours are included, what dramatically reduces the number of partonic channels.  This simplification
allows us to keep the presentation rather compact, while already including a large set of infrared 
limits appearing at NNLO QCD. We compare the results obtained for the various cross-section
contributions, and in particular the achieved variances, for different phase-space sampling techniques, including the default approach used in \Stripper, and in particular the Normalizing-Flow methods introduced in Sec.~\ref{sec:flows}. We thereby consider training separate samplers for the positive and negative contributions to the integrands, requiring accurate 
predictions of the sign of the target function. As a second figure of merit, we consider the corresponding unweighting efficiencies that determine the resources needed to obtain unit-weight events.   

\subsection{Calculational setup}

We consider proton--proton collisions at $\sqrt{s}=13\,\text{TeV}$, corresponding to the conditions of Run II at the LHC. For the proton PDFs the \textsc{NNPDF31\_nnlo\_as\_0118} set~\cite{NNPDF:2017mvq} is used. We consider the final-state top quarks to be on-shell using $m_t = 173.3\,\text{GeV}$. The mass parameter is thereby renormalized on-shell. The renormalization ($\mu_R$) and factorization ($\mu_F$) scales we set to the same fixed value
\begin{equation}
\mu_R=\mu_F = m_t/2\,.
\end{equation}
The renormalization of the strong coupling is performed in the $5$-ﬂavour scheme, assuming as an input value $\alpha_s(M^2_Z)=0.118$, consistent with the PDF set. 

\begin{figure}
    \centering
    \includegraphics[width=0.7\textwidth]{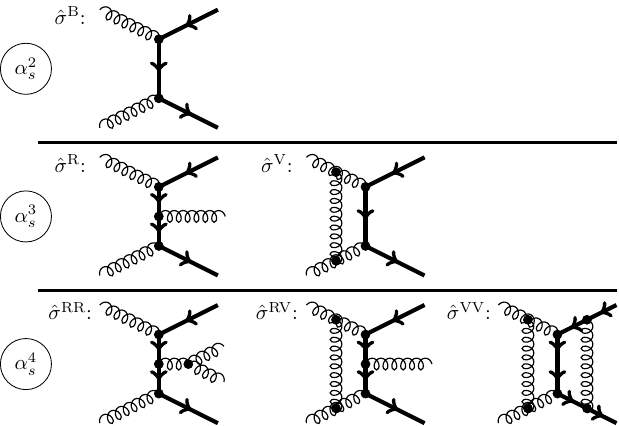}
    \caption{Representative Feynman diagrams contributing to gluonic top-quark pair production at NNLO QCD.}
    \label{fig:feyn-top-quarks}
\end{figure}

For illustration, in Fig.~\ref{fig:feyn-top-quarks} we present example Feynman diagrams contributing to gluonic top-quark pair production at LO $\propto {\cal{O}}(\alpha_s^2)$, NLO $\propto {\cal{O}}(\alpha_s^3)$, and NNLO $\propto {\cal{O}}(\alpha_s^4)$, thereby indicating to which \Stripper contribution they belong to, cf.\ Eqs.~\eqref{eq:sigmaR}--\eqref{eq:sigmaVV}. Note, for the here considered scale setting, i.e.\ $\mu_R=\mu_F$, the contributions $\sigma^{\text{C}}$, $\sigma^{\text{C1SU}}$, and $\sigma^{\text{C2DU}}$ vanish. Furthermore, $\sigma^{\text{VU}}$, $\sigma^{\text{VVDU}}$, and $\sigma^{\text{VVFR}}$ generally contribute to $\epsilon$-poles only. Accordingly, we omit these contributions in the following discussion. Also, these contributions are negligible in terms of their computational complexity and thus do not significantly affect our computational efficiency considerations.

\subsection{Integrators and optimization}\label{sec:flows_for_nnlo}

In Sec.~\ref{subsec:integrands}, we already discussed the structure of the integrands and the phase-space parameterization in the \Stripper formalism and how they can be cast into integrals over unit-hypercubes, see Eq.~\eqref{eq:PSI_general}. Apart from the discrete variable $l$, these integrals are precisely in the form of Eq.~\eqref{eq:integral}.

In practice, we developed a Python interface to the \Stripper framework, and the sampling and training procedures are steered through a Python script. From the interface point of view, a random point from the unit-hypercube with a weight (either uniform or transformed through the integrator) needs to be provided. This point is passed to the \Stripper C++ code, which returns the full event weight. In the case of the flow-based models, we treat the index $l$ as a conditional parameter to the neural networks we train, as explained in Sec.~\ref{subsec:conditional_flows}. During sampling, we use precomputed weights, based on the contribution to the total cross section of the integral with index $l$, to sample the index $l$. In the case of the \Vegas integrator, this is achieved by treating $l$ as an additional discretized dimension.

Apart from providing the unit-hypercube random samples, we want to stress that the integrand is handled entirely within the \Stripper framework. This allows access to all established facilities to compute differential cross sections, scale variations and unweighting for the \textsc{HighTEA} \cite{Czakon:2023hls} database.

We study the variance-reduction capability of three different adaptive phase-space samplers when applied to gluonic $t\bar{t}$ production in the \Stripper framework:
\begin{itemize}
    \item \textsc{Vegas}: The \textsc{Vegas} importance-sampling algorithm~\cite{Lepage:1977sw} as implemented in the Python package~\cite{pyvegas_url} is used. 100 channels, i.e.\ discrete bins, are used per phase-space dimension. This integrator has been cross-checked against a one-dimensional version of the \textsc{Parni} algorithm~\cite{vanHameren:2007pt}, which presents the default for phase-space integration within the \Stripper framework for NNLO QCD calculations. Since the performance of these two integrators was evaluated to be practically identical, we only show results for \textsc{Vegas} in this work.
    \item \textsc{Cl}: A Normalizing Flow sampler based on Coupling Layers as described in Sec.~\ref{sec:coupling_layer_flows} is used. We employ the \textsc{Nflows} package~\cite{nflows:2020} to implement piecewise rational-quadratic Coupling Layers~\cite{Durkan:2019} with 16 spline knots per dimension. The number of Coupling Layers is given by \(2 \lceil \log_2 n \rceil\), where \(n\) is the number of phase-space dimensions. Between two Coupling Layers, the inputs are permuted with a binary permutation scheme. Each coupling transformation is parameterized by a multilayer perceptron (MLP) with a structure that depends on the number of phase-space dimensions, which determines the number of input and output variables. The number and size of hidden layers is also variable. All hidden layers have twice as many nodes as the preceding layer, so that their size is increasing exponentially. The largest hidden layer has more nodes than the output layer, but less than twice as many. In our examples, this results in MLPs with up to 3 layers and flows with $\sim 100$k to $\sim 20$M trainable parameters. We use ReLU activation functions for the hidden layers. The discrete parameter $l$ is incorporated as a conditional parameter, as described in Sec.~\ref{subsec:conditional_flows}. It is transformed into a continuous representation by an embedding layer, where the number of nodes is equal to the number of possible values of \(l\). We concatenate this representation to the other input variables in each Coupling Layer.
    \item \textsc{Ode}: Corresponds to the Continuous Normalizing Flow sampling as described in Sec.~\ref{sec:ode_flows}. For training the time-dependent vector field, we use the implementation of the Flow-Matching method in the \textsc{TorchCFM} package~\cite{tong2024improving,tong2023simulation}. The vector field is parameterized by an MLP with four hidden layers of width 512, using SELU activation functions~\cite{NIPS2017_5d44ee6f}. This results in models with roughly 1M trainable parameters. The time variable \(t\) is concatenated to the other input variables. The discrete index \(l\) is one-hot encoded and then also concatenated to the other inputs. During training, the time \(t\) is sampled from a power-law distribution \(p_\alpha(t) \propto t^{1/(1+\alpha)},\ t \in [0,1]\), as proposed in~\cite{NEURIPS2023_3663ae53}. We set \(\alpha=1\). For the parameter \(\sigma\) appearing in Eq.~\eqref{eq:CFM_conditional_density}, we use a value of \(\sigma=\num{1e-4}\). For using the vector field in a Continuous Normalizing Flow and to sample from the flow, we use the \textsc{TorchDYN} package~\cite{politorchdyn}. The sampling requires to numerically solve an ODE, for which we use the Dormand--Prince method~\cite{DORMAND198019} of order 5 and set the absolute and relative tolerances to \num{1e-4}.
\end{itemize}
The quoted settings (e.g.\ number of Coupling Layers, number of neural-network parameters, or channels) result from some simple and not too-extensive hyperparameter optimization. 

Both types of flow are implemented and trained with the use of the \textsc{PyTorch} library~\cite{NEURIPS2019_9015}. We use the \textsc{AdamW} optimizer~\cite{DBLP:conf/iclr/LoshchilovH19} with a learning rate of \num{1e-3}. For the \textsc{Cl} flow, we reduce the learning rate on plateaus. To facilitate an efficient training of the flow\-/based integrators on GPUs, we perform the training iteratively. For the first training iteration, we generate a uniformly distributed sample $\vec{y}$ including the weight and discrete selection variable $l$, and write it to disk.

The training procedure depends on the flow model. For the \textsc{Cl} flow, the data is split into a training (80\%) and validation set (20\%), which we found necessary to avoid over-fitting. We train for up to 600 epochs while we allow it to stop if a plateau in the validation loss is detected. Each epoch is split into mini-batches based on randomly permuted subsets of the training data with a size of about 10k events. In the case of the \textsc{Ode} flow, we train on the complete data set for 600 epochs with a mini-batch size of about 100k. The large mini-batch size reduces the variance of the estimator for the flow-matching loss, Eq.~\eqref{eq:cfm_objective}, leading to faster convergence. Finally, the trained model is used to generate a new set of phase-space points $\vec{y}$ and prior weights. The integrand is evaluated at these points, and the weights are written to disk, which enables the next training iteration. At each iteration, we load the previous iteration of the model as the starting point and determine new weights for the categorical distribution of the index \(l\).

How much a given model (with fixed hyperparameters) can reduce the variance depends on two main aspects: 1) how much training data is used and 2) the expressivity of the parameterization. The first aspect is rather evident, but the second is more subtle. Assuming an infinite amount of training data, it is clear that a given model might not be able to describe the integrand arbitrarily well. This is because the parameterization might not be able to capture the actual functional form or correlations between variables (most evident for \textsc{Vegas}).

We consider the same amount of training data to compare different models, i.e.\ the same number of phase-space points, which we choose to be $10^7$. However, the training protocols depend on the model. The \textsc{Vegas} methods are updated every $10^4$ points while the flow-based models are trained on $10^6$ points in large batches. After the model is trained, all parameters are frozen, and the integrand is evaluated with $10^6$ points. These samples are then used to investigate the variance and convergence.

\subsection{Results}

We performed the above training procedures on all integrals contributing to gluonic top-quark pair production at NNLO QCD, both, on the absolute and the positive/negative split integrands. After the training, i.e.\ adaptation, of the three sampling methods, we performed the numerical integration with $10^6$ points and based our analyses on the corresponding event samples. To validate the integrators and benchmark their performance, we computed the following quantities:
\begin{itemize}
    \item $\hat{\sigma}$ and $\delta \hat{\sigma}$: the Monte Carlo estimate for the full hadronic cross section and its statistical uncertainty based on the integrator trained on the absolute value of the integrand.
    \item $\hat{\sigma}^{\pm}$ and $\delta \hat{\sigma}^{\pm}$: the Monte Carlo estimate and uncertainty for the positive and negative contribution to the full integral, respectively, based on the integrator separately trained on the positive/negative integrand.
    \item $\hat{\Sigma}^{\pm}$ and $\delta \hat{\Sigma}^{\pm}$: The sum of $\hat{\sigma}^+$ and $\hat{\sigma}^-$, the uncertainty is thereby rescaled to reflect a sample size of $10^6$ points in total. The ratio of effective points for the positive contribution, cf.\ Eq.~\eqref{eq:alpha_introduced}, is given by: $$ \alpha = \frac{\sqrt{\operatorname{Var}(\hat\sigma^+)}}{\sqrt{\operatorname{Var}(\hat\sigma^+)}+\sqrt{\operatorname{Var}(\hat\sigma^-)}}\,.$$
    \item $\epsilon^{\pm}_{\Phi}$: the efficiency to correctly hit the positively/negatively contributing phase-space region $\Phi^\pm$ in case of positive/negative split integrand.
    \item $\epsilon^{\pm}$: the unweighting efficiency for the positive and negative contribution, respectively.
    \item $\epsilon^{\pm}_{0.1\%}$: the unweighting efficiency after the reduction of the respective $w_{\rm max}$ based on removing the largest weights that contribute less than $0.1\%$ of the corresponding integrated cross section, see Sec.~\ref{sec:prelim1} for details.
\end{itemize}
The results for the various contributions are collected in Tab.~\ref{tab:num_res_nlo} in App.~\ref{app:results} for the Born and NLO terms and in Tabs.~\ref{tab:num_res_nnlo_RR} and \ref{tab:num_res_nnlo_V} for the double-real and virtual NNLO corrections, respectively. Alongside, we present an estimate for the CPU cost for a single point (relative to the evaluation of a $\sigma^{\rm B}$ point) and the estimated optimal uncertainty $\delta_{\rm opt}$ when training on the absolute integrand. All integral estimates are statistically compatible within a few multiples of their respective statistical uncertainty, a critical validation criterion.

\subsubsection*{Integral-variance reduction: training on the absolute integrand}

The prime performance characteristics of an integrator are its expressiveness and adaptability to the integrand, which should be reflected in the variance of the weights and, therefore, in the statistical uncertainty of the integral estimate, when considering a fixed number of integrand evaluations.

We first discuss the canonical approach, in which the integrators are trained on the absolute value of the integrand and the entire phase space is integrated together. For integrands of non-definite sign, if no positive/negative splitting is performed, there is a lower bound on the uncertainty, denoted as $\delta_{\text{opt}}$, based on the positive-to-negative ratio, see Eq.~\eqref{eq:minerr_toy}. Suppose the adaptation of the integrator, i.e.\ its sampling distribution, to the integrand is expressive enough and there is a sufficiently balanced positive-to-negative ratio. In that case, this effect will dominate the estimated uncertainty, and there is no improvement to be expected beyond that point by the sheer expressiveness of the integrator. In the case of a negligible negative (or positive, for that matter) contribution, as for instance in the Born cross section $\sigma^{\rm B}$ arising from negative PDF values, however, the reduction of the variance of the positive (or negative) part dominates. The expected enhanced expressiveness of the Normalizing Flow-based methods can be seen in the corresponding uncertainty estimates. For the Born contribution, the \Vegas integrator results in a cross-section estimate of $\hat{\sigma}^{\rm B}_{\text{\Vegas}}=532.8\pm 0.2\,\text{pb}$, while  for the flow samplers we obtain $\hat{\sigma}^{\rm B}_{\textsc{Ode}}=532.79\pm 0.05\,\text{pb}$ and $\hat{\sigma}^{\rm B}_{\textsc{Cl}}=532.80\pm 0.03\,\text{pb}$, respectively. Accordingly, the uncertainty gets reduced by a factor $4$ for the \textsc{Ode} flow and a factor $7$ for the \textsc{Cl} flow, corresponding to a performance gain of a factor $16$ and $49$, respectively. However, the formal bound on the uncertainty of $\delta_{\text{opt}}=0.01\,\text{pb}$ is not yet reached. A smaller but still visible performance gain can be observed in more complicated contributions with subtractions like $\sigma^{\rm RU}$, $\sigma^{\rm RRF}$, $\sigma^{\rm RRSU}$ and $\sigma^{\rm RVF}$, where we find a reduction of the uncertainties by a factor of $2$ with respect to the {\sc{Vegas}} method if a flow-based integrator is used. The performances of the \textsc{Ode} and \textsc{Cl} samplers are thereby largely on par. Importantly, we can observe that both integrators in these cases saturate or are close to saturating the corresponding uncertainty bound $\delta_{\rm opt}$. Further, we find that for all the other (simpler) NLO and NNLO QCD contributions ($\sigma^i\;\text{for}\; i \in \{ {\rm RF,\rm VF, RVFR, RVDU, VVF, RRDU} \}$), this bound can be saturated by all integrators. For example, for $\hat\sigma^{\rm RF}$ the bound evaluates to $\delta_{\text{opt}}=0.8\,\text{pb}$, and we find uncertainties for the three samplers given by $\delta \hat{\sigma}^{\rm RF}_{\text{\Vegas}} = 1\,\text{pb}$, and $\delta \hat{\sigma}^{\rm RF}_{\textsc{Ode}} = \delta \hat{\sigma}^{\rm RF}_{\textsc{Cl}} = 0.9\,\text{pb}$, respectively.

\begin{figure}[t]
    \centering
    \includegraphics[width=0.95\linewidth]{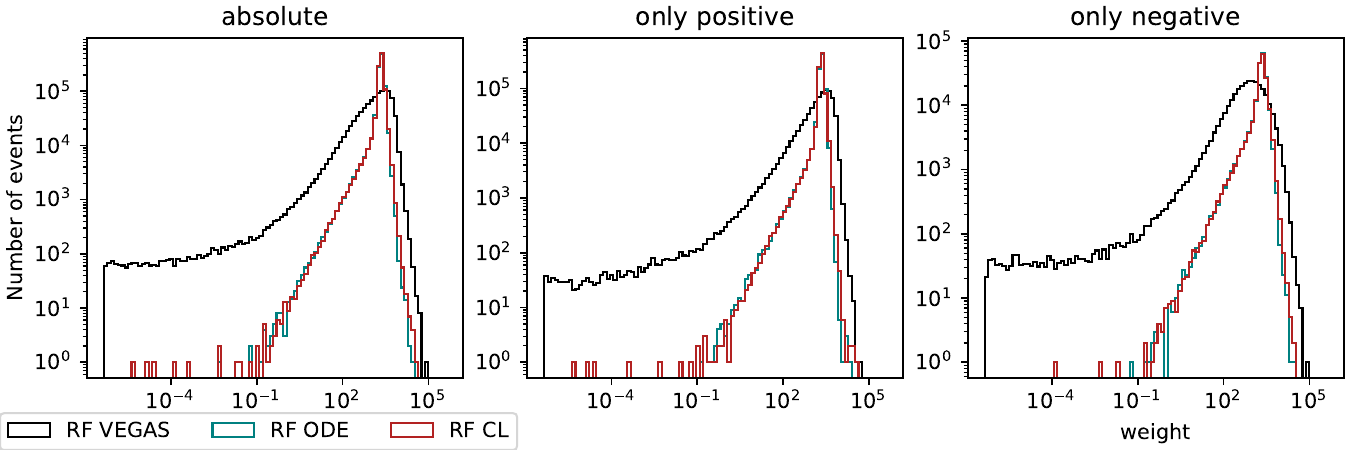}
    \includegraphics[width=0.95\linewidth]{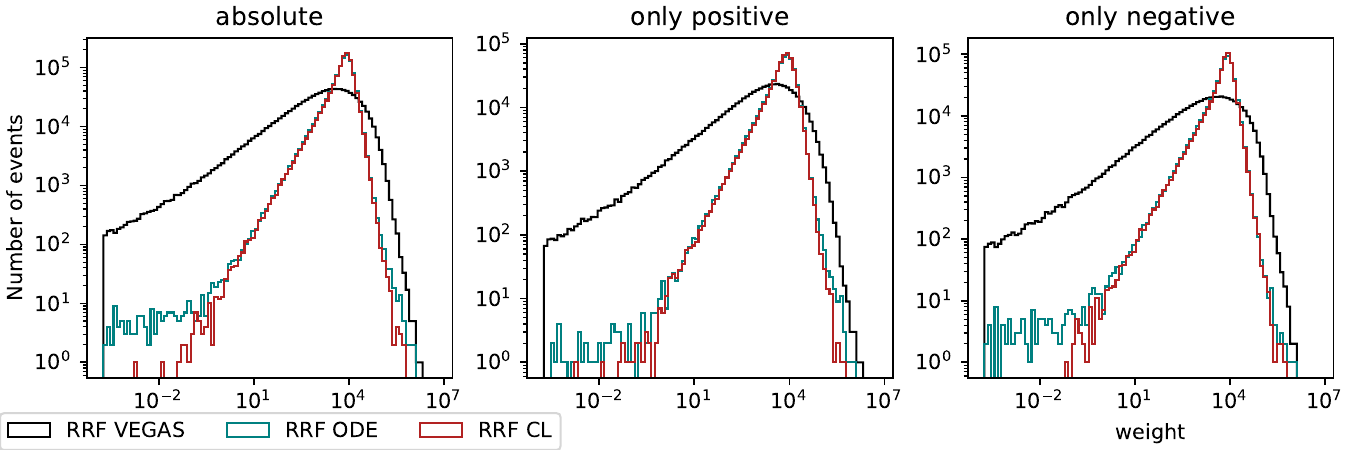}
    \caption{The distribution of weights for the $\sigma^{\rm RF}$ (top row) and $\sigma^{\rm RRF}$ contributions (bottom row). The first column shows the absolute weight, and the second and third columns show the positive and negative weight distributions. The distributions are shown for the \textsc{Vegas}, \textsc{Cl} and \text{ODE} integrators, all trained on the absolute integrand.}
    \label{fig:example-weight-distribution}
\end{figure}

Besides the uncertainty of the integral estimate, the actual weight distribution can give insights into how well an integrator learned the target distribution. For $\sigma^{\rm RF}$ and $\sigma^{\rm RRF}$, we show the weight distribution for the absolute integrand as well as the distributions of the positive and negative weights separately in Fig.\ \ref{fig:example-weight-distribution}. First, it can be noticed that the distributions of the \textsc{Ode} and \textsc{Cl} integrators are much more narrow than the \Vegas ones, and second, we can see that the maximum weights are reduced, i.e.\ there are significantly less events with large weights in the \textsc{Ode} and \textsc{Cl} samples. The distributions suggest that the performance gain using the flow-based integrators should be significant. As mentioned, the only mild performance gain in these cases is caused by sizable cancellations between the positive and negative contributions.

\begin{figure}
    \centering
    \includegraphics[width=0.49\linewidth]{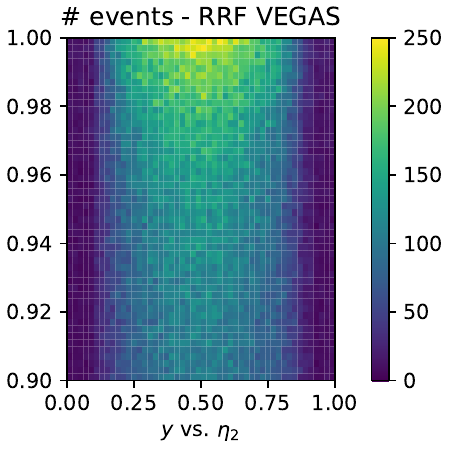}%
    \hfill
    \includegraphics[width=0.49\linewidth]{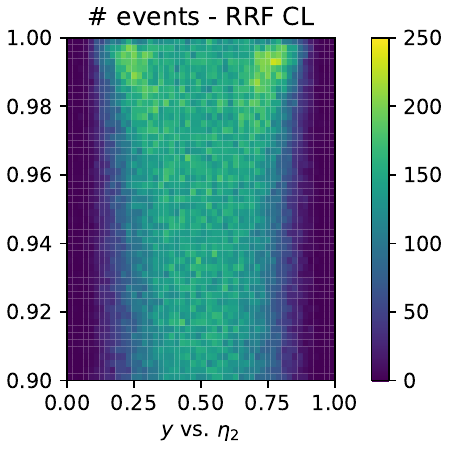}
    \caption{Example of non-factorizable correlations between two phase-space parameters for the $\sigma^{\rm RRF}$ contribution. The left panel shows the learned sampling distribution of \textsc{Vegas}, and the right panel that of the \textsc{Cl} integrator.}
    \label{fig:example-2d-non-factor}
\end{figure}

The reduction in the weight variance can be traced back to non-factorizable features in the phase space that the flow-based methods can learn, but the \Vegas model can not. A visualization of such a feature is given in Fig.~\ref{fig:example-2d-non-factor} for the case of the $\sigma^{\rm RRF}$ contribution. We show a 2D histogram (the counts are indicated through the heat map), with the coordinate axes given by two integration parameters, for the \Vegas integrator and the \textsc{Cl} integrator (the \textsc{Ode} one looks identical to the \textsc{Cl} one). The non-trivial structure learned by the \textsc{Cl} integrator can be seen by eye, and it is clear that \Vegas learned a different distribution.

\subsubsection*{Integral-variance reduction: training on split integrand}

\begin{figure}[t]
    \centering
    \includegraphics[width=0.49\linewidth]{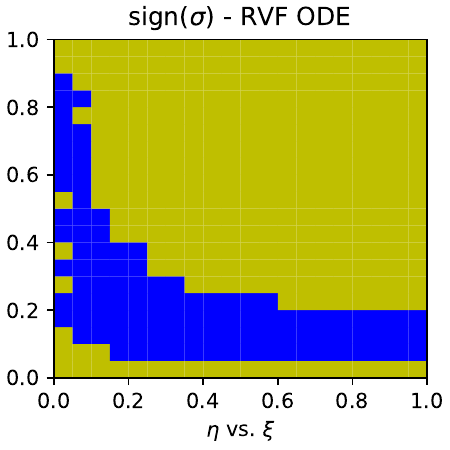}
    \caption{The sign (positive - yellow, negative - blue) of the binned cross section of the $\sigma^{\rm RVF}$ contribution as a function of two input parameters is shown. All other dimensions are integrated out.}
    \label{fig:2-sign}
\end{figure}

The advantage of using a flow-based method is clearly diminished when there are large cancellations between positive and negative contributions, and the gain based on variance reduction is thus limited. This implies that further improvement on the quality of the integrator, through more parameters or training data, will not lead to significant performance improvements. The picture changes, however, when a positive/negative split integrand is considered and the positive and negative regions are treated as independent integrals. In this case, there is no lower bound on the variance upon integration, i.e.\ a vanishing uncertainty can in principle be achieved. Therefore, the improved adaptability of the flows can be fully exploited. However, the integration problem is now more challenging because the efficiency of hitting the right phase-space region becomes crucial for reaching a small variance (missed points are counted as zero weight, increasing the variance). The phase-space boundaries between the positive and negative regions are non-trivial and likely non-factorizable. In Fig.~\ref{fig:2-sign} we show the sign of the integral, differential and binned in two integration variables, for the $\sigma^{\rm RVF}$ contribution, as an example for such non-trivial boundaries. This is only a two-dimensional projection and in reality the boundaries between the signs have a complicated shape in multidimensional space. Naturally, we expect the \Vegas algorithm to have trouble performing well in this split integrand case.

The results for the positive and negative cross-section contributions $\hat\sigma^\pm$ individually are given in Tabs.~\ref{tab:num_res_nlo}, \ref{tab:num_res_nnlo_RR} and \ref{tab:num_res_nnlo_V}, together with their combination $\hat\Sigma^{\pm}$. The correctness of this approach is validated by the agreement of $\hat\Sigma^{\pm}$ with the canonical approach, i.e.\ the respective $\hat\sigma$, within the estimated statistical uncertainty. To gauge the performance gain of the positive/negative splitting, the uncertainty of this quantity has to be compared to $\delta \hat\sigma$. First, a general observation: we find an improved or at least comparable performance for all integrators compared to the traditional non-split integrands with the same integrator. We can sometimes observe significant performance gains when using the flow-based integrators. A relevant example at NLO is the $\sigma^{\rm RF}$ contribution. Both, for the \textsc{Ode} and the \textsc{Cl} sampler, the uncertainty drops by a factor of $3$ from $\delta\hat\sigma^{\text{RF}}=0.9\,\text{pb}$, obtained when training on the absolute integrand, to $\delta\hat\Sigma^{\text{RF}}=0.3\,\text{pb}$ based on the split integrand. This corresponds to a factor of $9$ in performance gain. In fact, both samplers manage to clearly go below the uncertainty limit of $\delta_{\text{opt}}=0.8\,\text{pb}$ when training on the absolute integrand. Such improvement is absent for the \Vegas integrator, what can be traced back to its poor phase-space efficiency $\epsilon^{\pm}_{\Phi}$, which only reaches $73\%$ and $34\%$ for the positive and negative phase-space region, respectively. The flow-based methods instead reach efficiencies of about $99\%$ in this case. At NNLO QCD, the most impressive case is $\sigma^{\rm RVF}$, which features the highest CPU evaluation costs per phase-space point. For the RVF contribution the \textsc{Ode} sampler reduces the uncertainty by a factor $3$ upon positive/negative splitting. However, the rather challenging RRF contribution does not benefit from the splitting, given the training setup, as sampling efficiencies $\epsilon^\pm_{\Phi}$ of only $85\%$ are reached.

Eventually, not only the achieved gain for each cross-section contribution is relevant, but it is also crucial to factor in the corresponding computational costs and the relative contribution to the final result. To this end, the amount of phase-space points evaluated for each individual contribution needs to be optimized. This can be realized for a requested target uncertainty given by
\begin{equation}
    \delta \sigma_{\rm target} = \delta \hat\sigma_{\rm tot} (\vec{n}) = \sqrt{\sum_i (\delta \hat\sigma^i(n_i))^2} = \sqrt{\sum_i \frac{\operatorname{Var}(\hat\sigma^i)}{n_i-1}}\,,
\end{equation}
where the sum $i$ runs over all contributions included, $\operatorname{Var}(\hat\sigma^i)$ denotes the corresponding weight variance, and the elements $n_i$ of vector $\vec{n}$ represent the number of points in each contribution. Under this constraint the computational costs $C$, defined by 
\begin{equation}
    C = \sum_i C_i n_i\,,
\end{equation}
can then be minimized, with $C_i$ the costs per phase-space point for contribution $i$. This minimization problem can be solved numerically using Lagrange multipliers. Folding the costs and contributions to the final result in this way, we are left with a gain in computational efficiency of a factor $\approx 2$ for the flow-based methods in the case of training on the absolute integrand and a factor $\approx 8$ in the case of positive/negative separated integrands with respect to our \Vegas baseline. At NNLO QCD accuracy, the calculation is dominated ($\approx 80\%$) by the computation of the double-real contributions, followed by the real-virtual part. Note, the above cost estimate does not include the GPU time for training and inference of the phase-space sampler. This implies that the quoted gains present an upper bound, reached in the limit where the matrix-element evaluation time clearly dominates.

For reference, for the $RRF$ contribution (which has the largest flows and hence NNs) we found that we need 3 (6) minutes for one training iteration and 1 (3) minutes for the generation step on one half of a Nvidia A100 chip for the \textsc{Cl}(\textsc{Ode})-flow. These numbers imply a total training and generation time for this case of approximately 30 to 90 minutes for the entire training sequence. These are to be compared with approximately 10 hours of integrand evaluation time on a CPU, totalling 10 million evaluations. However, the GPU numbers should be taken with a grain of salt, as I/O operations contribute significantly due to the short runtime.

\begin{figure}[t]
    \centering
    \includegraphics[width=0.6\linewidth]{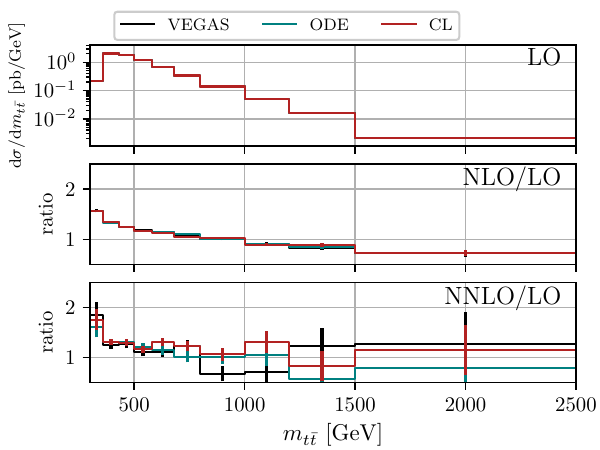}
    \caption{Differential cross section with respect to the top-pair invariant mass $m_{t\bar{t}}$. The upper panel shows the absolute LO QCD result obtained with the \Vegas, \textsc{ODE} and \textsc{Cl} integrator. The middle and lower panels show the NLO and NNLO QCD results as a ratio to the LO result, respectively. The vertical bars indicate the statistical uncertainty.}
    \label{fig:ttx-diff-mtt-abs}
\end{figure}

\subsubsection*{Differential cross sections}

\begin{figure}[t]
    \centering
    \includegraphics[width=0.5\linewidth]{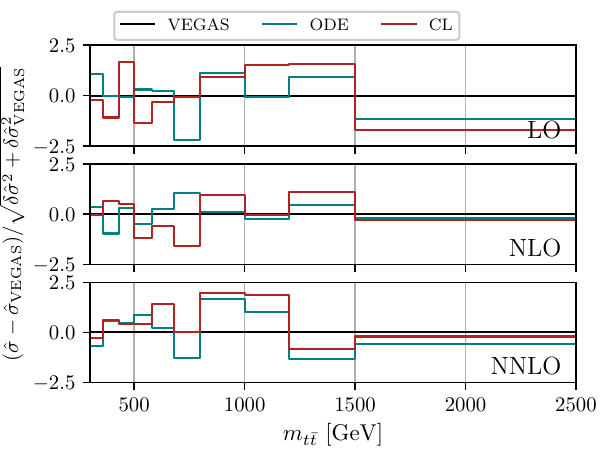}%
    \includegraphics[width=0.5\linewidth]{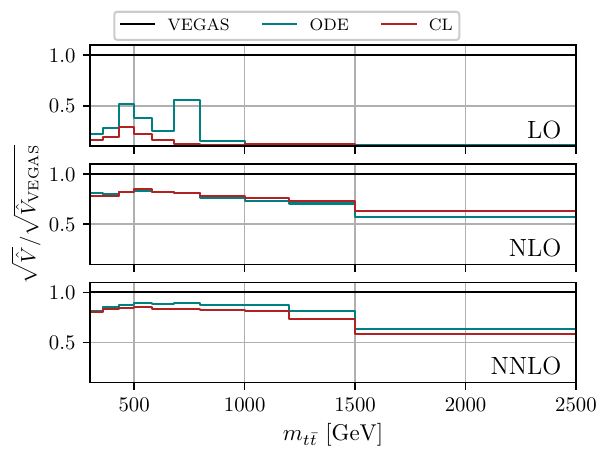}
    \caption{The left-hand column shows the agreement between \textsc{Vegas}, \textsc{Ode} and \textsc{Cl} as a function of the top-pair invariant mass $m_{t\bar{t}}$. The right-hand side shows the variance of the weights in each bin as a ratio to the \Vegas variance.}
    \label{fig:ttx-diff-mtt}
\end{figure}

Up to here, we focused on the total cross section, which is also the central training objective for our integrators. However, we want to explicitly stress that we can simultaneously compute arbitrary differential cross sections for infrared and collinear safe observables with our methods. As an example, in Fig.~\ref{fig:ttx-diff-mtt-abs} we show the differential distribution with respect to the invariant mass of the top-quark pair at LO, as obtained with the three considered integration algorithms. The corresponding NLO and NNLO results are presented as ratios to the LO. Notably, the three integrators faithfully reproduce the differential distributions, and the different results agree within their statistical uncertainty in each bin, which is highlighted on the left-hand side of Fig.~\ref{fig:ttx-diff-mtt}, showing the deviation relative to the combined statistical uncertainty. Further, on the right-hand side in Fig.~\ref{fig:ttx-diff-mtt}, we can also demonstrate that the flow-based integrators reduce the weight variance in each bin with respect to the \Vegas method.

\subsubsection*{Unweighted event generation}

The capability to reduce the variance of the sample weights is not the only figure of merit that can be considered when comparing importance-sampling methods. A different metric is provided by the unweighting efficiency, given by $\epsilon = \langle w \rangle / w_{\rm max}$. Accordingly, the unweighting efficiency depends on the weight maximum and is thus rather sensitive to large-weight outliers. In Tabs.~\ref{tab:num_res_nlo}, \ref{tab:num_res_nnlo_RR} and \ref{tab:num_res_nnlo_V} we quote the achieved unweighting efficiencies for the positive and negative region of phase space as $\epsilon^+$ and $\epsilon^-$ separately. In addition, we provide the efficiency estimate when considering a reduced maximum weight. This reduced maximum weight is determined such that the largest weights contributing cumulatively less than $0.1\%$ to the total cross section are discarded, cf.\ Sec.~\ref{sec:prelim1} for details. This quantity, denoted as $\epsilon^\pm_{0.1\%}$, is less sensitive to weight fluctuations within a fixed sample and leads to higher efficiencies by design. We want to note that if the largest-weight event corresponds to a substantial fluctuation, i.e.\ a single event contributing more than $0.1\%$ to the total integral, then no improvement can be achieved with this prescription. The improvements in terms of computational costs are linearly correlated with the efficiency.

We can observe increases in the unweighting efficiency, specifically in the $\epsilon^{\pm}_{0.1\%}$ measure, using the flow-based integrators in various contributions. The unweighting efficiency for contributions with tree-level kinematics, such as $\sigma^{\rm B}$, $\sigma^{\rm VF}$ or $\sigma^{\rm VVF}$, improve by a factor of $2$ to $4$ with respect to \Vegas when using the flow-based approaches. Further, we see similar improvements for single-radiation contributions. For example, $\sigma^{\rm RF}$ and  $\sigma^{\rm RVF}$, see an improvement by a factor of $3$ in unweighting efficiency using the \textsc{Ode} flow compared to \Vegas. However, for $\sigma^\text{RVF}$ the \textsc{Cl} sampler yields a rather low efficiency of $\epsilon^+_\textsc{Cl}=7.6\cdot 10^{-4}$, that is dominated by a large-weight outlier, given that $\epsilon^+_{0.1\%,\textsc{Cl}}$ has the same value. For the \textsc{Ode} sampler we find for the nominal maxima efficiencies of $\epsilon^+_{\textsc{Ode}}=2.9\cdot 10^{-3}$ and $\epsilon^-_{\textsc{Ode}}=5.1\cdot 10^{-2}$, compared to the \Vegas values of $\epsilon^+_{\text{\Vegas}}=1.3\cdot 10^{-2}$ and $\epsilon^-_{\text{\Vegas}}=1.4\cdot 10^{-2}$. After reducing the maxima, however, the \textsc{Ode} sampler outperforms \Vegas for both regions, where $\epsilon^+_{0.1\%,\textsc{Ode}}=6.7\cdot 10^{-2}$ vs. $\epsilon^+_{0.1\%,\text{\Vegas}}=2.4\cdot 10^{-2}$ and $\epsilon^-_{0.1\%,\textsc{Ode}}=2.5\cdot 10^{-1}$ vs. $\epsilon^-_{0.1\%,\text{\Vegas}}=3.8\cdot 10^{-2}$. For the double-real contributions, i.e.\ $\sigma^{\rm RRF}$, $\sigma^{\rm RRSU}$ and $\sigma^{\rm RRDU}$, we do not find significant gains with respect to \Vegas.

These observations can be correlated with the weight distributions shown in Fig.~\ref{fig:example-weight-distribution}. We see that while the flow-based methods feature visibly sharper peaks, both, for the positive and negative parts, the maximum weights (i.e.\ the ends of the tail to the right) are similar to the \Vegas results, in particular for the $\sigma^{\rm RRF}$ contribution. While the outlined prescription for reducing the maximum weight does lead to similar performances of \Vegas and the flow-based models, it is clear that the number of events with overweights will be substantially larger in the \Vegas case. This results in a smaller effective sample size compared to the samples generated with the flow methods.

While \Stripper is not an event generator, (partially) unweighted events can be used to store a calculation on disk for further analysis in the \textsc{HighTEA} framework \cite{Czakon:2023hls}. The gains in unweighting efficiency directly translate into reduced computational costs for the event generation.

\section{Conclusions}\label{sec:conclusions}

In this work we studied the application of two Normalizing-Flow integrators for the evaluation of NNLO QCD cross sections. In particular, we considered Coupling Layer and Continuous Normalizing Flows and applied them to top-quark pair production in the gluonic channel. The NNLO QCD calculation follows the \Stripper approach for the subtraction of soft and collinear singularities. In order to reach NNLO QCD accuracy, besides the lowest-order Born contribution, various (subtracted) single- and double real-emission and one- and two-loop virtual corrections, as well as interference terms need to be evaluated, and several contributing partonic channels have to be considered. The \Stripper subtraction formalism is based on selector functions that address specific soft and collinear limits and, consequently, sectorize phase space. In turn, this results in a significant proliferation of phase-space integrals that need to be performed. Furthermore, the integrands of the various contributions are often not of definite sign, what limits the performance of any integration method whose sampling distribution is adapted to the absolute value of the integrand, as is canonically done. To address this issue, we consider the stratification of the integrands into their positive and negative parts and separately train the sampling algorithms. Accordingly, for an optimal sampling distribution the integral variance would vanish. However, for this approach to be successful the sampler has to efficiently predict the sign of the integrand for a given phase-space point. Furthermore, we developed the means to condition the flow-based samplers on other discrete variables, including a sector index and helicity states. 

As main figure of merit for our novel sampling methods we consider the statistical uncertainty of the obtained cross-section estimates. Furthermore, we monitor the resulting weight distributions and from these extract unweighting efficiencies for performing a rejection sampling to obtain unit-weight events. To this end, we studied the impact of a systematic reduction of the maximum weights, corresponding to the production of partially unweighted events, but with potentially increased efficiency. We compare all our results against the \Vegas sampler, that closely reproduces the performance of the current default method used in \Stripper, namely the \textsc{Parni} algorithm~\cite{vanHameren:2007pt}.

The enhanced expressiveness and adaptability of the flow-based samplers to the target distribution with respect to the factorized density learned by \Vegas results in significant reductions of the weight-distribution variances. In particular when splitting the integrands into their positive and negative contributions did we find large performance gains for both neural importance samplers. Overall this can be summarized in the computational costs for obtaining an estimate for the total cross section at predefined statistical uncertainty. When dominated by the time needed to evaluate the integrands, the costs can be reduced by a factor of 2 when training the samplers on the absolute integrand, and a factor of 8 when splitting the integrands into their positive and negative parts. Besides for the total cross sections, the flow-based samplers also yield reduced weight variances in binned differential distributions, as exemplified for the top-pair invariant mass observable. For the unweighting efficiencies of the various cross-section contributions we also find improvements when using flow-based samplers, although rare large-weight outliers can sometimes obscure this picture. 

An improved unweighting efficiency is rather a byproduct of the variance\-/reduction training incentive of importance samplers. Large-weight outliers, that impair the unweighting efficiency, will often originate from rarely sampled phase-space regions,  where the samplers have not seen a lot of training data. To improve on that, the phase-space distribution could be biased towards overproducing such rare configurations. This procedure could also be employed to reduce the statistical uncertainty in the tails of differential distributions. 

In summary, the application of modern flow-based sampling methods provides large potential to significantly reduce the computational costs of cutting edge higher-order calculations. We have worked out a rather simple workflow for training and application of the samplers, showcased for top-quark pair production at NNLO QCD with \Stripper, that could similarly be used for other calculational frameworks. 

\section*{Acknowledgements}
We acknowledge fruitful discussions with David Yallup at an early stage of the project. TJ would like to thank Bernhard Schmitzer and Fabian Sinz for insightful discussions about Continuous Normalizing Flows and Flow Matching. TJ and SS are grateful for financial support from BMBF (projects 05D23MG1 and 05H24MGA) and funding by the Deutsche Forschungsgemeinschaft (DFG, German Research Foundation) through project 456104544.  TJ expresses his 
gratitude to CIDAS for supporting him with a fellowship.

The authors gratefully acknowledge the computing time granted by the Resource Allocation Board and provided on the supercomputer Emmy/Grete at NHR-Nord@Göttingen as part of the NHR infrastructure. The calculations for this research were conducted with computing resources under the project nhr\_ni\_starter\_22045.

RP acknowledges the support of Narodowe Centrum Nauki under Grant No. 2023/49/B/ST2/04330 (SNAIL). The authors acknowledge support from the COMETA COST Action CA22130.

\clearpage

\appendix

\section{Numerical results}\label{app:results}

\SetTblrInner{
    rowsep=1pt,
    colsep=4pt
}
\begin{table}[h]
    \sisetup{text-series-to-math = true, 
        detect-weight=true,
        mode=text,
    }
    \begin{footnotesize}
    \begin{center}
    \begin{tblr}[evaluate=\fileInput]{
        hline{1,Z} = {\heavyrulewidth},
        hline{2,4} = {\lightrulewidth},
        vline{2}={23,26}{solid,abovepos = -1,belowpos = -1},
        vline{5}={22,25}{solid,abovepos = -1},
        vline{5}={23,26}{solid},
        vline{5}={24,27}{solid,belowpos = -1},
        vline{7}={22,25}{solid,abovepos = -1},
        vline{7}={23,26}{solid},
        vline{7}={24,27}{solid,belowpos = -1},
        vline{9}={22,25}{solid,abovepos = -1},
        vline{9}={23,26}{solid},
        vline{9}={24,27}{solid,belowpos = -1},
        vline{11}={22,25}{solid,abovepos = -1},
        vline{11}={23,26}{solid},
        vline{11}={24,27}{solid,belowpos = -1},
        colspec = {@{}lllS[table-format=+1.4(1)e1,table-column-width=3.5em]S[table-format=1.3]S[table-format=+1.4(1),table-column-width=4em]S[table-format=1.3]S[table-format=+1.4(1),table-column-width=4em]S[table-format=1.3]S[table-format=+1.5(1),table-column-width=3.5em]S[table-format=1.2]@{}},
        row{1} = {guard},
        cell{1,2,3}{1} = {c=3}{}, 
        cell{4,7,10,13,16,19,22,25}{1} = {r=3}{},
        cell{22,25}{2} = {r=3}{},
        cell{4,7,10,13,16,19}{1} = {c=2}{},
        cell{1}{4,6,8,10} = {c=2}{c}, 
        cell{2-21}{4} = {c=2}{l}, 
        cell{2-21}{6} = {c=2}{l},
        cell{2-21}{8} = {c=2}{l},
        cell{2-21}{10} = {c=2}{l},
        column{5} = {rightsep=20pt},
        cell{6}{4} = {font=\bftab}, 
        cell{5,6}{6} = {font=\bftab},
        cell{5,6}{8} = {font=\bftab},
        cell{4,5,6}{10} = {font=\bftab},
        cell{9}{4} = {font=\bftab}, 
        cell{8,9}{6} = {font=\bftab},
        cell{8,9}{8} = {font=\bftab},
        cell{9}{10} = {font=\bftab},
        cell{12}{4} = {font=\bftab}, 
        cell{11}{6} = {font=\bftab},
        cell{12}{8} = {font=\bftab},
        cell{12}{10} = {font=\bftab},
        cell{15}{4} = {font=\bftab}, 
        cell{14,15}{6} = {font=\bftab},
        cell{14,15}{8} = {font=\bftab},
        cell{15}{10} = {font=\bftab},
        cell{17,18}{4} = {font=\bftab}, 
        cell{17,18}{6} = {font=\bftab},
        cell{17,18}{8} = {font=\bftab},
        cell{17,18}{10} = {font=\bftab},
        cell{20,21}{4} = {font=\bftab}, 
        cell{20,21}{6} = {font=\bftab},
        cell{20,21}{8} = {font=\bftab},
        cell{20,21}{10} = {font=\bftab},
        cell{24}{4} = {font=\bftab}, 
        cell{24}{5} = {font=\bftab},
        cell{22}{6} = {font=\bftab},
        cell{23}{7} = {font=\bftab},
        cell{22}{8} = {font=\bftab},
        cell{23}{9} = {font=\bftab},
        cell{24}{10} = {font=\bftab},
        cell{24}{11} = {font=\bftab},
        cell{27}{4} = {font=\bftab}, 
        cell{27}{5} = {font=\bftab},
        cell{26}{6} = {font=\bftab},
        cell{26}{7} = {font=\bftab},
        cell{25}{8} = {font=\bftab},
        cell{27}{9} = {font=\bftab},
        cell{27}{10} = {font=\bftab},
        cell{27}{11} = {font=\bftab},
        }
        \fileInput{table-performance-NLO.tex}
    \end{tblr}
    \end{center}
    \end{footnotesize}
  \caption{Results obtained with the \Vegas, \textsc{Ode} and \textsc{Cl} integrators for the Born and NLO QCD contributions to gluonic top-quark pair production. The training of all samplers is based on 
$10^7$ points ($10 \times 10^6$ points with intermediate adaptation). The quoted results are based on $10^6$ integrand evaluations with frozen integrator settings. Note, for better readability all cross-section values have been rescaled by factors indicated in the first row, while all efficiencies are
quoted with their nominal value.}
  \label{tab:num_res_nlo}
\end{table}

\cleardoublepage

\begin{table}[h]
    \sisetup{text-series-to-math = true, 
        detect-weight=true,
        mode=text,
    }
    \begin{footnotesize}
    \begin{center}
    \begin{tblr}[evaluate=\fileInput]{
        hline{1,Z} = {\heavyrulewidth},
        hline{2,4} = {\lightrulewidth},
        vline{2}={23,26}{solid,abovepos = -1,belowpos = -1},
        vline{5}={22,25}{solid,abovepos = -1},
        vline{5}={23,26}{solid},
        vline{5}={24,27}{solid,belowpos = -1},
        vline{7}={22,25}{solid,abovepos = -1},
        vline{7}={23,26}{solid},
        vline{7}={24,27}{solid,belowpos = -1},
        vline{9}={22,25}{solid,abovepos = -1},
        vline{9}={23,26}{solid},
        vline{9}={24,27}{solid,belowpos = -1},
        colspec = {@{}lllS[table-format=+1.4(1),table-column-width=4em]S[table-format=1.4]S[table-format=+1.5(1),table-column-width=4.5em]S[table-format=1.5]S[table-format=+1.4(1),table-column-width=4em]S[table-format=1.4]@{}},
        row{1} = {guard},
        cell{1,2,3}{1} = {c=3}{}, 
        cell{4,7,10,13,16,19,22,25}{1} = {r=3}{},
        cell{22,25}{2} = {r=3}{},
        cell{4,7,10,13,16,19}{1} = {c=2}{},
        cell{1}{4,6,8} = {c=2}{c}, 
        cell{2-21}{4} = {c=2}{l}, 
        cell{2-21}{6} = {c=2}{l},
        cell{2-21}{8} = {c=2}{l},
        cell{5,6}{4} = {font=\bftab}, 
        cell{6}{6} = {font=\bftab},
        cell{6}{8} = {font=\bftab},
        cell{9}{4} = {font=\bftab}, 
        cell{8,9}{6} = {font=\bftab},
        cell{8,9}{8} = {font=\bftab},
        cell{11,12}{4} = {font=\bftab}, 
        cell{10,12}{6} = {font=\bftab},
        cell{12}{8} = {font=\bftab},
        cell{14,15}{4} = {font=\bftab}, 
        cell{15}{6} = {font=\bftab},
        cell{15}{8} = {font=\bftab},
        cell{18}{4} = {font=\bftab}, 
        cell{17}{6} = {font=\bftab},
        cell{17}{8} = {font=\bftab},
        cell{20,21}{4} = {font=\bftab}, 
        cell{21}{6} = {font=\bftab},
        cell{21}{8} = {font=\bftab},
        cell{22}{4} = {font=\bftab}, 
        cell{24}{5} = {font=\bftab},
        cell{24}{6} = {font=\bftab},
        cell{23,24}{7} = {font=\bftab},
        cell{23}{8} = {font=\bftab},
        cell{23}{9} = {font=\bftab},
        cell{25}{4} = {font=\bftab}, 
        cell{27}{5} = {font=\bftab},
        cell{25}{6} = {font=\bftab},
        cell{25}{7} = {font=\bftab},
        cell{27}{8} = {font=\bftab},
        cell{25}{9} = {font=\bftab},
        }
        \fileInput{table-performance-NNLO-RR.tex}
    \end{tblr}
    \end{center}
    \end{footnotesize}
    \caption{Results obtained with the \Vegas, \textsc{Ode} and \textsc{Cl} integrators for the double-real NNLO QCD contributions to gluonic top-quark pair production. The training of all samplers is based on 
$10^7$ points ($10 \times 10^6$ points with intermediate adaptation). The quoted results are based on $10^6$ integrand evaluations with frozen integrator settings. Note, for better readability all cross-section values have been rescaled by factors indicated in the first line, while all efficiencies are
quoted with their nominal value. }
    \label{tab:num_res_nnlo_RR}
\end{table}

\begin{table}[ht]
    \sisetup{text-series-to-math = true, 
        detect-weight=true,
        mode=text,
    }
    \begin{footnotesize}
    \begin{center}
    \begin{tblr}[evaluate=\fileInput]{
        hline{1,Z} = {\heavyrulewidth},
        hline{2,4} = {\lightrulewidth},
        vline{2}={23,26}{solid,abovepos = -1,belowpos = -1},
        vline{5}={22,25}{solid,abovepos = -1},
        vline{5}={23,26}{solid},
        vline{5}={24,27}{solid,belowpos = -1},
        vline{7}={22,25}{solid,abovepos = -1},
        vline{7}={23,26}{solid},
        vline{7}={24,27}{solid,belowpos = -1},
        vline{9}={22,25}{solid,abovepos = -1},
        vline{9}={23,26}{solid},
        vline{9}={24,27}{solid,belowpos = -1},
        vline{11}={22,25}{solid,abovepos = -1},
        vline{11}={23,26}{solid},
        vline{11}={24,27}{solid,belowpos = -1},
        colspec = {@{}lllS[table-format=+1.5(1),table-column-width=4.5em]S[table-format=1.5]S[table-format=+1.4(1),table-column-width=4em]S[table-format=1.3]S[table-format=+1.4(1),table-column-width=3.5em]S[table-format=1.2]S[table-format=+1.4(1),table-column-width=4em]S[table-format=1.2]@{}},
        row{1} = {guard},
        cell{1,2,3}{1} = {c=3}{}, 
        cell{4,7,10,13,16,19,22,25}{1} = {r=3}{},
        cell{22,25}{2} = {r=3}{},
        cell{4,7,10,13,16,19}{1} = {c=2}{},
        cell{1}{4,6,8} = {c=2}{c}, 
        cell{2-21}{4} = {c=2}{l}, 
        cell{2-21}{6} = {c=2}{l},
        cell{2-21}{8} = {c=2}{l},
        cell{2-21}{10} = {c=2}{l},
        cell{5,6}{4} = {font=\bftab}, 
        cell{4,5,6}{6} = {font=\bftab},
        cell{4,5,6}{8} = {font=\bftab},
        cell{4,5,6}{10} = {font=\bftab},
        cell{8}{4} = {font=\bftab}, 
        cell{8}{6} = {font=\bftab},
        cell{8,9}{8} = {font=\bftab},
        cell{9}{10} = {font=\bftab},
        cell{11}{4} = {font=\bftab}, 
        cell{11}{6} = {font=\bftab},
        cell{12}{8} = {font=\bftab},
        cell{12}{10} = {font=\bftab},
        cell{14}{4} = {font=\bftab}, 
        cell{14}{6} = {font=\bftab},
        cell{15}{8} = {font=\bftab},
        cell{15}{10} = {font=\bftab},
        cell{17,18}{4} = {font=\bftab}, 
        cell{17}{6} = {font=\bftab},
        cell{17,18}{8} = {font=\bftab},
        cell{17,18}{10} = {font=\bftab},
        cell{20,21}{4} = {font=\bftab}, 
        cell{21}{6} = {font=\bftab},
        cell{20,21}{8} = {font=\bftab},
        cell{20,21}{10} = {font=\bftab},
        cell{22}{4} = {font=\bftab}, 
        cell{23}{5} = {font=\bftab},
        cell{22}{6} = {font=\bftab},
        cell{23}{7} = {font=\bftab},
        cell{23}{8} = {font=\bftab},
        cell{23}{9} = {font=\bftab},
        cell{24}{10} = {font=\bftab},
        cell{23}{11} = {font=\bftab},
        cell{26}{4} = {font=\bftab}, 
        cell{26}{5} = {font=\bftab},
        cell{25}{6} = {font=\bftab},
        cell{26}{7} = {font=\bftab},
        cell{27}{8} = {font=\bftab},
        cell{27}{9} = {font=\bftab},
        cell{26}{10} = {font=\bftab},
        cell{27}{11} = {font=\bftab},
        }
        \fileInput{table-performance-NNLO-V.tex}
    \end{tblr}
    \end{center}
    \end{footnotesize}
    \caption{Results obtained with the \Vegas, \textsc{Ode} and \textsc{Cl} integrators for the virtual NNLO QCD contributions to gluonic top-quark pair production. The training of all samplers is based on 
$10^7$ points ($10 \times 10^6$ points with intermediate adaptation). The quoted results are based on $10^6$ integrand evaluations with frozen integrator settings. Note, for better readability all cross-section values have been rescaled by factors indicated in the first line, while all efficiencies are
quoted with their nominal value. }
    \label{tab:num_res_nnlo_V}
\end{table}

\clearpage

\bibliographystyle{JHEPmod}
\bibliography{lit}

\end{document}